\documentclass[fleqn,usenatbib]{mnras}
\usepackage{newtxtext,newtxmath}
\usepackage[T1]{fontenc}

\usepackage{amsmath}
\usepackage{mathtools}
\usepackage{gensymb}
\usepackage[english]{babel}
\usepackage{lineno}
\usepackage{lscape}
\usepackage{soul}
\usepackage{graphicx}
\usepackage[dvipsnames]{xcolor}
\usepackage{bm}
\usepackage{epstopdf}
\usepackage{url}
\usepackage{mdframed}
\usepackage{booktabs}
\usepackage[super]{nth}
\usepackage[math]{cellspace}

\usepackage{fontawesome}
\usepackage{microtype}
\usepackage{xspace}
\usepackage{verbatim}
\usepackage{epigraph}
\usepackage{import}
\usepackage[noabbrev, capitalize]{cleveref}
\usepackage{arydshln}
\usepackage{pifont}
\newcommand{\xmark}{\ding{55}}

\newcommand\vect[1]{\boldsymbol{#1}}

\newcommand\ii{\mathrm{i}}

\newcommand\e[1]{_{\mathrm{#1}}}
\newcommand\h[1]{^{\mathrm{#1}}}

\newcommand{\dd}{\mathrm{d}}

\newcommand{\ddf}[3][]{\frac{\dd^{#1} #2}{\dd {#3}^{#1}}}

\newcommand{\IPhT}{%
Universit\'{e} Paris-Saclay, CNRS, CEA, Institut de physique th\'{e}orique, 91191, Gif-sur-Yvette, France}

\newcommand{\INAF}{%
INAF - Osservatorio Astronomico di Roma, via Frascati 33, 00040 Monteporzio Catone (Roma), Italy}

\newcommand{\Montpellier}{%
Laboratoire Univers et Particules de Montpellier, Universit\'{e} de Montpellier, CNRS, Montpellier, France}

\newcommand{\INFN}{INFN - Sezione di Roma, Piazzale Aldo Moro, 2 - c/o Dipartimento di Fisica, Edificio G. Marconi, I-00185 Roma, Italy}

\usepackage[authoryear]{natbib}
\usepackage{hyperref}

\DeclareRobustCommand{\VAN}[3]{#2}

\crefname{figure}{Fig.}{Figs.}
\Crefname{figure}{Figure}{Figures}

\newcommand{\amplification}{\bm{\mathcal{A}}}
\newcommand{\Sersic}{S\'{e}rsic\xspace}
\newcommand{\mSersic}{S\acute{e}rsic\xspace}
\newcommand{\lens}{\texttt{lenstronomy}\xspace}
\newcommand{\productive}{\mathcal{P}}
\newcommand{\excess}{>2\sigma}
\newcommand{\minmodel}{minimal model\xspace}
\newcommand{\linfit}{$\log_{10}\sigma\approx -0.76 \log_{10} Q + 1.01$\xspace}

\newcommand{\modification}[1]{#1}

\title[Measuring LOS shear with Einstein rings]{Measuring line-of-sight shear with Einstein rings: a proof of concept}

\author[N.~B.~Hogg et al.]{
Natalie~B.~Hogg,$^{1}$\thanks{E-mail: natalie.hogg@ipht.fr} 
Pierre~Fleury,$^{1}$ 
Julien~Larena,$^{2}$ 
and Matteo~Martinelli$^{3,4}$ 
\\
$^{1}$\IPhT\\
$^{2}$\Montpellier\\
$^{3}$\INAF\\
$^{4}$\INFN
}

\date{Accepted 2023 February 13. Received 2023 February 13; in original form 2022 October 21}

\pubyear{2023}

\begin{document}
\label{firstpage}
\pagerange{\pageref{firstpage}--\pageref{lastpage}}
\maketitle

\begin{abstract}
Line-of-sight effects in strong gravitational lensing have long been treated as a nuisance. However, it was recently proposed that the line-of-sight shear could be a cosmological observable in its own right, if it is not degenerate with lens model parameters. We firstly demonstrate that the line-of-sight shear can be accurately measured from a simple simulated strong lensing image with percent precision. We then extend our analysis to more complex simulated images and stress test the recovery of the line-of-sight shear when using deficient fitting models, finding that it escapes from degeneracies with lens model parameters, albeit at the expense of the precision. Lastly, we check the validity of the tidal approximation by simulating and fitting an image generated in the presence of many line-of-sight dark matter haloes, finding that an explicit violation of the tidal approximation does not necessarily prevent one from measuring the line-of-sight shear.
\end{abstract}

\begin{keywords} 
gravitational lensing: strong -- cosmology: theory -- software: development
\end{keywords}

\section{Introduction}

Strong gravitational lensing is one of the most elegant probes of the late-time Universe. Massive objects such as galaxies and galaxy clusters bend the local spacetime geometry and produce multiple, distorted images of distant background sources. When the source--lens--observer alignment is good, the image forms an Einstein ring. In an otherwise perfectly homogeneous and isotropic Universe, the morphology of that ring is a direct probe of the gravitational field generated by the strong lens.

However, on the scales probed by strong gravitational lensing, the Universe is very inhomogeneous. Therefore, we expect that inhomogeneities -- typically dark matter haloes -- near the line of sight (LOS) will generate weak lensing distortions that will affect the images produced by the strong lensing process. We generically refer to the effects induced by this weak lensing of strong lensing images as \textit{LOS effects}. 
\modification{These effects have been studied for their impact on, for example, the determination of the Hubble parameter via time-delay cosmography \citep{Gilman_2020}, in which case they are considered to be a nuisance effect.} But when LOS effects are observable, they also provide some insight into the distribution of dark matter. For example, the effects of individual haloes of dark matter, whether in the main lens or along the LOS, can lead to the detection of dark matter structures on small scales  \citep{Vegetti_2009,Vegetti_2012,Despali:2017ksx,Seng_l_2022,Nightingale:2022bhh} and the study of their properties \citep{Vegetti_2009,Vegetti_2014,Ostdiek:2020mvo,Zhang:2022djp}. Line-of-sight haloes and main-lens subhaloes lead to different signatures on strong lensing images, and their effects might be disentangled \citep{Dhanasingham:2022nox}, although the detectability of individual haloes \modification{remains an open question}~\citep{He:2021rjd}. Finally, LOS effects in strong lensing may constitute a novel probe of weak lensing in their own right, which is the motivation of the present work.

Line-of-sight perturbations can be modelled in two complementary ways. One option is to add thin perturbers along the LOS in a multi-plane lensing formalism \citep{1986ApJ...310..568B}, a methodology which is particularly suitable for taking into account the presence of several subdominant yet clearly identifiable mass concentrations near the LOS. However, \modification{this approach is not suitable for modelling the collective effect of a large number of perturbers, due to an excessive number of extra parameters thereby added to the lens model.}

This difficulty can be overcome by encapsulating the collective effects of LOS inhomogeneities into an effective, external convergence and shear perturbing the LOS~\citep{1987ApJ...316...52K, 1996ApJ...468...17B, Schneider:1997bq, Birrer:2016xku, 2021CQGra..38h5002F}. Although this neglects higher-order effects such as flexion, this parametric method is usually considered good enough on large scales. An analysis of both approaches and how to connect them is presented in \cite{Fleury:2021tke}, where one can also find a discussion of higher-order effects.

The idea that the LOS shear can, in principle, be extracted from strong lensing images in galaxy lensing was proposed by \cite{Birrer:2016xku, Birrer:2017sge} as an alternative probe of cosmic shear. Based on this approach, a first attempt at correlating this LOS shear with standard weak lensing was made by \cite{Kuhn:2020wpy}.
However, the set-up developed in this programme relied on a non-optimal treatment of LOS effects that overlooked some degeneracies between foreground shear and the main lens modelling. It also employed a somewhat optimistic and simple model for the main lens.

In this paper, we re-analyse the possibility of measuring the LOS shear with Einstein rings generated by galaxy lensing using the \minmodel introduced by \cite{Fleury:2021tke}, in lieu of the parametrisation used by \cite{Birrer:2016xku}. This model fully accounts for degeneracies between the main lens and its LOS corrections, thereby providing an effective description that depends on only one additional complex parameter. This parameter, denoted $\gamma\e{LOS}$, is expected to be mostly independent of the main lens model. In our analysis, we take great care in studying the level of complexity required when modelling the main lens, showing that too simplistic a model can result in biases and a loss of accuracy in the reconstruction of the LOS shear.

We present the first demonstration that the LOS shear \textit{can} be observed using a series of simulated strong lensing images of increasing complexity. In particular, we show that the LOS shear is generally measurable without systematic biases, and that the next-order LOS effects, while present, do not introduce further biases. These results are obtained thanks to our modifications of the \lens software\footnote{\url{https://github.com/lenstronomy/lenstronomy}.} \citep{Birrer:2018xgm, Birrer2021}, to which we contributed a new subpackage, \texttt{LineOfSight}, \modification{included in the software from version 1.11.0}.

The paper is organised as follows: in \Cref{sec:advantage} we recapitulate the LOS theory and demonstrate how degeneracies between the LOS shears and other parameters are avoided in the so-called \minmodel; in \Cref{sec:measurability_shear}, we show how the LOS shear is systematically measurable; in \Cref{sec:validity_tidal_regime} we explore the validity of the tidal regime. In \Cref{sec:conclusion} we present our conclusions. In \Cref{appendix:lenstronomy} we describe our modifications to the strong lensing software \lens, while \Cref{appendix:composite_lens_parameters} contains details of the parameter choices used to create the mock image data analysed in this work.

Throughout this work, we employ the spatially flat $\Lambda$CDM cosmological model described by the \textit{Planck} 2018 data release \citep{Aghanim:2018eyx}, with $H_0 = 67.4$ kms$^{-1}$ Mpc$^{-1}$ and $\Omega_{\rm m} = 0.315$.

\section{The minimal line-of-sight model and its advantages}
\label{sec:advantage}

In this section, we review the theoretical underpinnings of LOS effects as presented in \cite{Fleury:2021tke} and demonstrate how the inherent degeneracies can be absorbed in the so-called \textit{\minmodel}, paving the way for the LOS shear to be measured.

\subsection{Tidal perturbations to a strong lens}

As a ray of light passes through the inhomogeneous Universe from source to observer, its path is perturbed by the presence of massive objects. In the simplest strong lensing situation, the deflection of light is assumed to be caused by a single lens. In this case, the whole problem is encapsulated in the lens equation,
\begin{align}
\bm{\beta} = \bm{\theta} -  \bm{\alpha}(\bm{\theta}),
\label{eq:lens_eqn_single_plane}
\end{align}
where $\bm{\beta}$ is the angular position of the source, $\bm{\theta}$ is the angular position of the observed image and $\bm{\alpha}(\bm{\theta})$ is the displacement angle, which is in turn determined by the gravitational potential associated with the surface density of the lens, projected along the LOS.

A more elaborate description can be made by including the effects of other nearby objects on the image. The most general approach to deal with this situation is the multi-plane lensing set-up~\citep{1986ApJ...310..568B}, in which each additional lens mass is considered to reside in its own plane along the LOS. However, the recursive nature of the multi-plane lens equation makes analytical treatments of complicated lensing situations intractable. 

A simpler approach is to treat the inhomogeneities along the LOS as tidal perturbations which induce a shear and convergence on the image produced by the main lens. This has the advantage of a far simpler theoretical description than the multi-plane case~\citep{1987ApJ...316...52K, 1994A&A...287..349S, 1996ApJ...468...17B, Schneider:1997bq, 2021CQGra..38h5002F}; see \cite{McCully:2013fga, Fleury:2021tke} for explicit derivations of the perturbed lens equation in the tidal regime, starting from the general multi-plane framework.

\begin{figure}
\centering
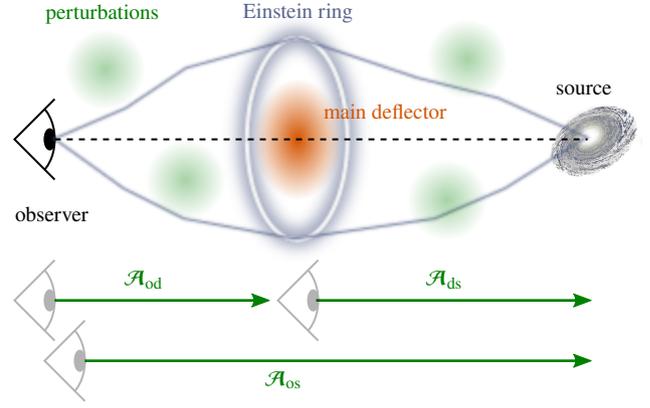
\caption{\modification{When LOS perturbations to a strong lens can be treated in the tidal regime, they intervene in the lens equation~\eqref{eq:lens_eqn_los} via three distortion matrices: $\amplification\e{od}$, $\amplification\e{os}$, $\amplification\e{ds}$. For each couple ab, $\amplification\e{ab}$ represents the distortions to a small source placed at (b) as observed from (a) due to the perturbers only.}}
\label{fig:perturbed_ring}
\end{figure}

In this simplified description, the lens equation~\eqref{eq:lens_eqn_single_plane} becomes
\begin{equation}
\bm{\beta}
= \amplification\e{os} \bm{\theta}
- \amplification\e{ds} \,
    \bm{\alpha}(\amplification\e{od} \bm{\theta}),
\label{eq:lens_eqn_los}
\end{equation}
where the amplification matrices~$\amplification\e{os}, \amplification\e{ds}, \amplification\e{od}$ encapsulate all the LOS perturbations. They are commonly parameterised as
\begin{equation}
\label{eq:decomposition_amplification_matrix}
\amplification\e{ab} = \bm{1} - \bm{\Gamma}\e{ab} \ ,
\quad
\bm{\Gamma}\e{ab} =
\begin{bmatrix}
\kappa\e{ab} + \rm{Re}\left(\gamma\e{ab}\right) & \rm{Im}\left(\gamma\e{ab}\right) - \omega\e{ab} \\ \rm{Im}\left(\gamma\e{ab}\right) + \omega\e{ab} & \kappa\e{ab}- \rm{Re}\left(\gamma\e{ab}\right)
\end{bmatrix},
\end{equation}
for $\mathrm{ab}\in\{\mathrm{os}, \mathrm{ds}, \mathrm{od}\}$.The quantities $\kappa\e{ab}, \gamma\e{ab}, \omega\e{ab}$ represent the convergence, the complex shear, and the rotation of an image as seen from the position (a) with respect to its source at the position (b), due to the inhomogeneities between (a) and (b) -- see \cref{fig:perturbed_ring}. For example, if a circular source were placed in the source plane (s), and observed from the main-lens plane (d), then $\gamma\e{ds}$ would be the complex ellipticity of its image due to the perturbations between (d) and (s). When all the perturbations are set to zero, $\bm{\Gamma}\e{os}=\bm{\Gamma}\e{ds}=\bm{\Gamma}\e{od}=\bm{0}$, we recover the standard single-plane lens equation~\eqref{eq:lens_eqn_single_plane}.

It is expected that the tidal regime, and hence \cref{eq:lens_eqn_los}, is valid when all the perturbers besides the main lens are either a smoothly distributed mass on the LOS, which would induce convergence, or compact masses lying far from the LOS, which would induce shear; rotation is caused by successive shears along different axes. Throughout this paper, we shall assume that $\kappa, \gamma, \omega$ are all very small, although \cref{eq:lens_eqn_los} would remain valid even for large values of those parameters. This description of a perturbed strong lensing system is simple enough to be easily implemented in and rigorously tested by the current state-of-the art strong lensing software, as we will describe in \Cref{sec:advantage} and \Cref{sec:measurability_shear}. It is, in addition, easily extended beyond the tidal regime following the dominant-lens approach of \cite{Fleury:2021tke}, as we will touch on in \Cref{sec:validity_tidal_regime}. 

\subsection{Absorbing degeneracies in a \minmodel}
\label{subsec:minimal_model}

In this work, we aim to show that the LOS shear can be measured from strong lensing images. However, as shown by \cite{Fleury:2021tke}, the three shears involved in \cref{eq:lens_eqn_los}, $\gamma\e{os}, \gamma\e{ds}, \gamma\e{od}$, cannot be measured independently, because they are degenerate with the main-lens properties and with one another. 
\modification{This degeneracy is a generalisation of the well-known mass-sheet degeneracy~\citep{Falco1985}, and is called the source-position transformation~\citep{Schneider:2013wga}, since it stems from our ignorance of the exact position and morphology of the source that is lensed.}
For example, one may multiply \cref{eq:lens_eqn_los} with any matrix $\amplification$ and redefine the unknown source position as $\bm{\beta}\to\amplification\bm{\beta}$; the resulting lens equation is formally identical to \cref{eq:lens_eqn_los}, but with $\amplification\e{os}\to\amplification\amplification\e{os}$ and $\amplification\e{ds}\to\amplification\amplification\e{ds}$.

Degeneracies may be removed by arbitrarily fixing a definition for the unknown source~$\bm{\beta}$. This is analogous to lifting a gauge freedom by fixing the gauge. As shown in \cite{Fleury:2021tke}, there exists a particularly meaningful choice, leading to a \emph{\minmodel} for the perturbed lens equation. The first step consists in rewriting the displacement angle as the gradient of a potential,
\begin{equation}
\bm{\alpha}(\bm{\theta})
= \ddf{\psi}{\bm{\theta}} \ ,
\end{equation}
where
\begin{equation}
\psi(\bm{\theta})
\equiv
\frac{D\e{ds}}{D\e{od}D\e{os}} \, \hat{\psi}(D\e{od}\bm{\theta}) \ ,
\end{equation}
and $\hat{\psi}(\bm{x})$ is twice the projected gravitational potential produced by the surface density of the main lens, $\Sigma(\bm{x})$,
\begin{equation}
\hat{\psi} (\bm{x})
\equiv
\frac{4G}{c^2} \int \dd^2 \bm{y} \; \Sigma(\bm{y}) \ln |\bm{x} - \bm{y}| \ ,
\end{equation}
where $G$ is Newton's constant and $c$ the speed of light.

Then, multiplying \cref{eq:lens_eqn_los} with the combination~$\amplification\e{od}\amplification\e{ds}^{-1}$, we obtain the \emph{\minmodel} for the lens equation,
\begin{equation}
\tilde{\bm{\beta}}
=
\amplification\e{LOS} \bm{\theta}
- \frac{\dd \psi_{\rm eff}}{\dd \bm{\theta}} \ ,
\label{eq:lens_eqn_minimal}
\end{equation}
with the transformed source position
$
\tilde{\bm{\beta}}
\equiv
\amplification\e{od}\amplification\e{ds}^{-1}\bm{\beta}
$,
and
\begin{align}
\amplification\e{LOS}
&\equiv
\amplification\e{od} \amplification\e{ds}^{-1} \amplification\e{os} \ ,
\label{eq:amplification_matrix_LOS}
\\
\psi\e{eff}(\bm{\theta})
&\equiv
\psi(\amplification\e{od}\bm{\theta}) \ .
\label{eq:effective_potential}
\end{align}
\Cref{eq:lens_eqn_minimal} effectively describes a main lens with potential $\psi\e{eff}$ and external tidal perturbations, $\amplification\e{LOS}$, located in the same plane.

Since \cref{eq:lens_eqn_los,eq:lens_eqn_minimal} are equivalent up to a source position transformation, the corresponding lens models would succeed or fail equally when fitting a strong lensing image. Therefore, any two distinct situations with the same $\amplification\e{LOS}$ and $\psi\e{eff}$ are observationally indistinguishable. This is why the (od), (os) and (ds) perturbations cannot be independently measured. However, when the perturbations are small, $|\bm{\Gamma}\e{od}|, |\bm{\Gamma}\e{ds}|, |\bm{\Gamma}\e{os}|\ll 1$, $\bm{\Gamma}\e{LOS}=\bm{1}-\amplification\e{LOS}$ is linearised as
\begin{equation}
\bm{\Gamma}\e{LOS}
\approx
\bm{\Gamma}\e{od} + \bm{\Gamma}\e{os} - \bm{\Gamma}\e{ds} \ .
\end{equation}
This means that the potentially measurable shear component is
\begin{equation}
\label{eq:LOS_shear}
\gamma\e{LOS} \approx \gamma\e{od} + \gamma\e{os} - \gamma\e{ds} \ .
\end{equation}
We shall see in \cref{subsec:simple_example} that $\gamma\e{LOS}$ can indeed be measured from a mock strong lensing image with a much higher precision than any one of $\gamma\e{od}, \gamma\e{os}$ or $\gamma\e{ds}$ taken independently.

Another degeneracy revealed by the \minmodel~\eqref{eq:lens_eqn_minimal} is between the foreground perturbations~(od) and the properties of the main lens, through the effective potential~$\psi\e{eff}$ of \cref{eq:effective_potential}. Specifically, the $\psi\e{eff}=\mathrm{cst}$ contours are the images of the $\psi=\mathrm{cst}$ contours by the foreground perturbations encoded in $\amplification\e{od}$. This generally makes it hard to distinguish between, for example, the effect of the ellipticity of the main lens and a foreground shear~$\gamma\e{od}$. Note, however, that in many widely used elliptical lens models, such as the singular isothermal ellipsoid~(SIE, \citealt{1994A&A...284..285K}), or its generalisation, the elliptical power law~(EPL, \citealt{2015A&A...580A..79T}), the foreground shear--ellipticity degeneracy is only approximate, because in those models the lens ellipticity is implemented at the level of the iso-density contours, not in the iso-potential contours. That is why, in their pioneering numerical experiments, \cite{Birrer:2016xku} could measure both $\gamma\e{od}$ and $\gamma\e{os}-\gamma\e{ds}=\gamma\e{LOS}-\gamma\e{od}$. Since this only applies to very specific lens models that are not necessarily well motivated from the astrophysical point of view, such an operation should nevertheless be considered very risky; we thus recommend not to try measuring $\gamma\e{od}$ in general, unless there is a compelling reason to trust the result.

We have chosen to focus the above discussion on the shear. The other components of the tidal matrix~$\bm{\Gamma}\e{LOS}$ are the convergence~$\kappa\e{LOS}$ and the rotation~$\omega\e{LOS}$. It is well known that the convergence and source position can be transformed in such a way that the observed lensing image is left invariant -- this is precisely the mass-sheet degeneracy -- which means that any attempt to measure or constrain the LOS convergence from a strong lensing image would be a pointless undertaking. For this reason, we shall keep the convergence parameters fixed to zero throughout this paper, both when generating mock images and when fitting them.

The rotation is induced by lens--lens coupling, and is generally small when the secondary lenses are individually weak, $\omega =\mathcal{O}(\gamma^2)$. However, it can play a significant role when the shear on the image is large -- say, $10\%$ or more. Shears this large would be unusual in real strong lensing data so when generating our mock images we restrict ourselves to smaller values, and keep the rotation parameters fixed to zero. Nevertheless, when fitting images with the \minmodel, we include $\omega\e{LOS}$ as a free (and in this case nuisance) parameter. Its presence accounts for any slight non-linear effect that may be induced by larger shears, and thus ensures a better recovery of $\gamma\e{LOS}$ from a given mock image.

\subsection{A simple example with \lens}
\label{subsec:simple_example}

To demonstrate the advantage of the \minmodel over the full model with a concrete example, we simulate a strong lensing image using the software \lens, which we modified to include the LOS formalism described above. Our modifications are now part of the public version of this code. For full details of our modifications of \lens, see \Cref{appendix:lenstronomy}. 

\begin{figure}
    \centering
    \includegraphics[width=0.8\columnwidth]{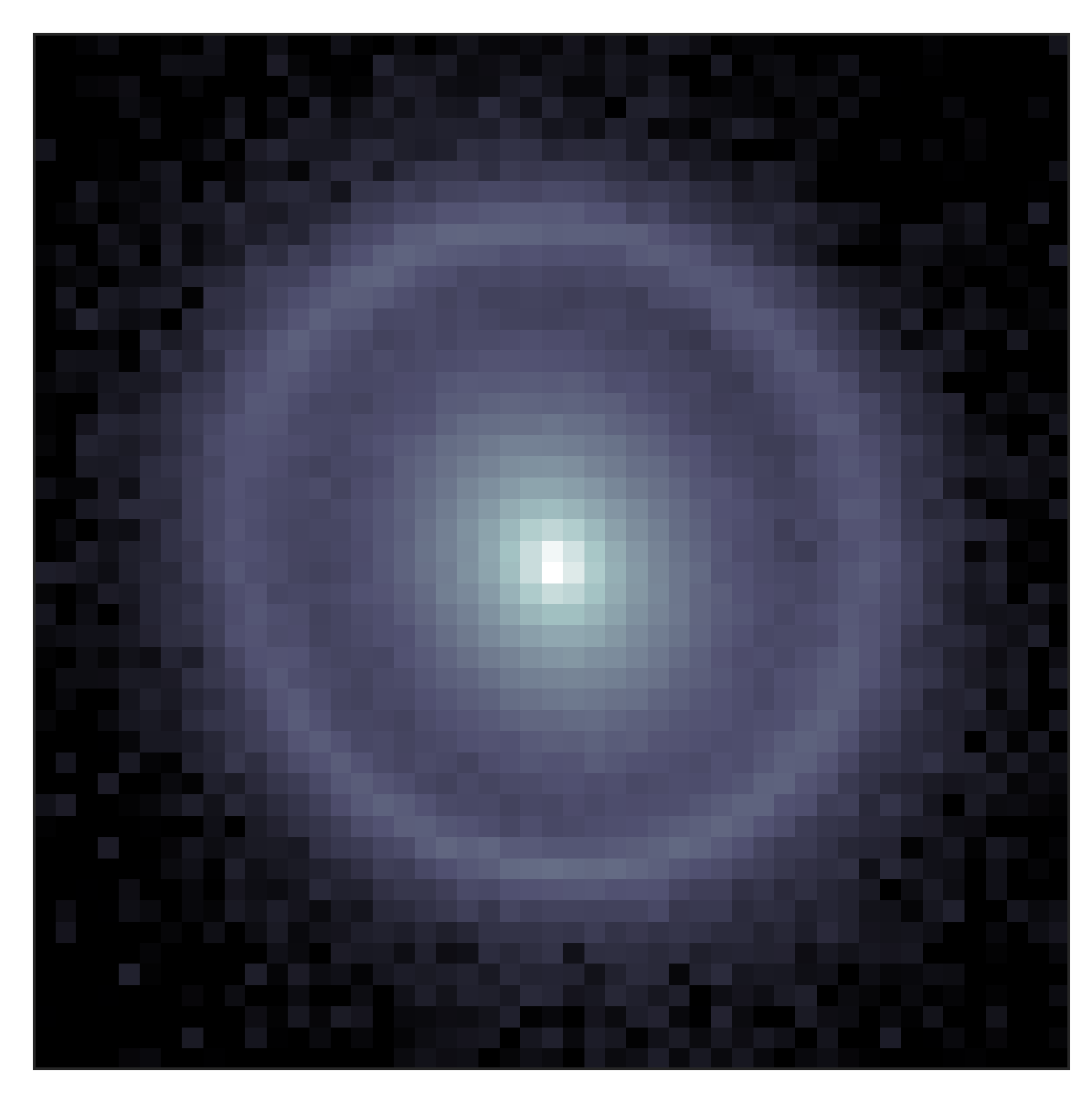}
    \caption{The image produced by a simple lens model consisting of an elliptical power law mass profile and a elliptical \Sersic lens light profile, with an elliptical \Sersic source (model parameters listed in \cref{tab:advantage_parameters}).}
    \label{fig:advantage_image}
\end{figure}

We create a simple lens using an EPL profile, with additional shears~$\gamma\e{od}, \gamma\e{os}, \gamma\e{ds}$. The ellipticity of the profile, along with its centre and the LOS shear components, are all drawn randomly. We model the source and the lens light with elliptical \Sersic light profiles. The ellipticity of the source is also generated at random.  Throughout this work we simulate the noise on our mock images according to the Hubble Space Telescope Wide Field Camera 3 F160W noise settings in \lens \citep{Windhorst2011}. The resulting simulated image is shown in \cref{fig:advantage_image} and the specific model parameters associated with this image are listed in \Cref{tab:advantage_parameters}.

\begin{table}
  \centering
  \caption{The parameters used to simulate the image shown in \cref{fig:advantage_image}, along with the priors on the parameters sampled in the MCMC to produce the results shown in \cref{fig:advantage_contours_full} and \cref{fig:advantage_contours_minimal}. The ellipticities, positions and shear parameters were all drawn at random. The lens light position and ellipticity are fixed to be the same as those of the EPL profile. The magnitudes quoted are apparent magnitudes. The expected values of the minimal LOS quantities $\gamma\e{LOS}, \omega\e{LOS}$ are derived from \cref{eq:amplification_matrix_LOS} using the decomposition of \cref{eq:decomposition_amplification_matrix}.}
  \begin{tabular}{SlSlSlSl}
    \hline
    \hline
    Component & Parameter  & Value & Prior \\
    \hline
                               & $\theta_{\rm E}$    &~ \modification{$1.2''$}  & $[0.01'', 10'']$ \\
                               & $\gamma$            &~ $2.6$   & $[0, 4.0]$ \\
Lens (EPL)                     & $e_1$               & $-0.053$  & $[-0.5, 0.5]$\\
                               & $e_2$               &~ $0.077$  & $[-0.5, 0.5]$\\
                               & $x$                 &~ $0.014''$ & -- \\
                               & $y$                 & $-0.048''$ & -- \\
                               \hline
                               & Magnitude           &~ $20$    & -- \\
Lens light (\Sersic ellipse)   & $R_{\rm{\mSersic}}$ &~ \modification{$0.5''$} & $[0.001, 10.0]$\\
                               & $n_{\rm{\mSersic}}$ &~ $4.0$   & $[0.5, 5.0]$\\
                               \hline
                               & Magnitude           &~ $24$      & -- \\
                               & $R_{\rm{\mSersic}}$ &~ \modification{$0.3''$}  & $[0'', 10'']$\\
                               & $n_{\rm{\mSersic}}$ &~ $1.0$     & $[0.5, 5]$\\
Source light (\Sersic ellipse) & $e_1$               &~ $0.104$   & $[-0.5, 0.5]$\\
                               & $e_2$               & $-0.038$   & $[-0.5, 0.5]$\\
                               & $x$                 & $-0.003''$ & $[-0.1'', 0.1'']$ \\
                               & $y$                 &~ $0.048''$ & $[-0.1'', 0.1'']$ \\
                               \hline
                               & $\gamma_1^{\rm od}$ & $-0.0027$ & $[-0.5, 0.5]$ \\
                               & $\gamma_2^{\rm od}$ & $-0.0060$ & $[-0.5, 0.5]$\\
LOS                            & $\gamma_1^{\rm os}$ & $-0.0091$ & $[-0.5, 0.5]$\\
                               & $\gamma_2^{\rm os}$ & $-0.028$  & $[-0.5, 0.5]$\\
                               & $\gamma_1^{\rm ds}$ &~ $0.019$  & $[-0.5, 0.5]$\\
                               & $\gamma_2^{\rm ds}$ &~ $0.018$  & $[-0.5, 0.5]$\\
                               \hdashline[2pt/3pt]
                               & $\gamma_1\h{LOS}$   & $-0.031$  & $[-0.5, 0.5]$\\
Derived                        & $\gamma_2\h{LOS}$   & $-0.052$  & $[-0.5, 0.5]$\\
                               & $\omega\e{LOS}$     & $0.00033$ & $[-0.5, 0.5]$\\
                               \hline
\end{tabular}

\label{tab:advantage_parameters}
\end{table}

Using the affine-invariant ensemble sampling Markov chain Monte Carlo (MCMC) parameter inference method available in the \texttt{emcee} package\footnote{\url{https://github.com/dfm/emcee}.} \citep{ForemanMackey2013}, we then fit the simulated image shown in \cref{fig:advantage_image} with two different models: a model comprised of an EPL plus the full LOS model~\eqref{eq:lens_eqn_los}, i.e. the same model as was used to generate the mock image; and a model comprised of an EPL plus the \minmodel~\eqref{eq:lens_eqn_minimal}. We sample all the parameters listed in \cref{tab:advantage_parameters} except for the lens and lens light central positions, which are kept fixed at the centre of the image. We impose uniform priors on all the parameters and run chains containing ten walkers per parameter. 
As discussed above, we expect that the individual shear components in the full model will be degenerate with each other, while the $\gamma\e{LOS}$ combination of the \minmodel will be accurately recovered. In both cases (full and \minmodel), we expect the foreground shear $\gamma\e{od}$ to be degenerate with the ellipticity of the main lens.

\begin{figure*}
    \centering
    \includegraphics[width=0.49\textwidth]{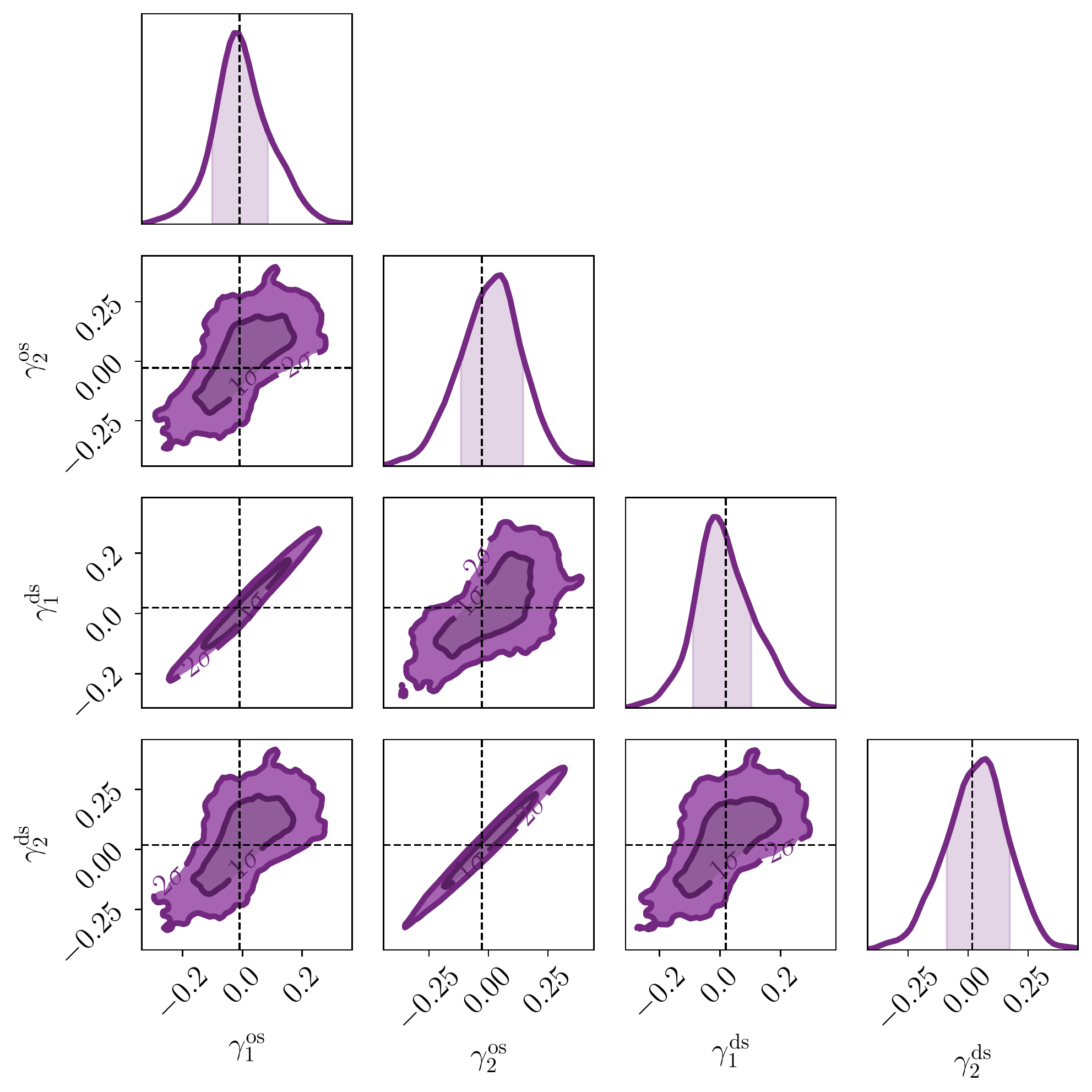}
    \hfill
    \includegraphics[width=0.49\textwidth]{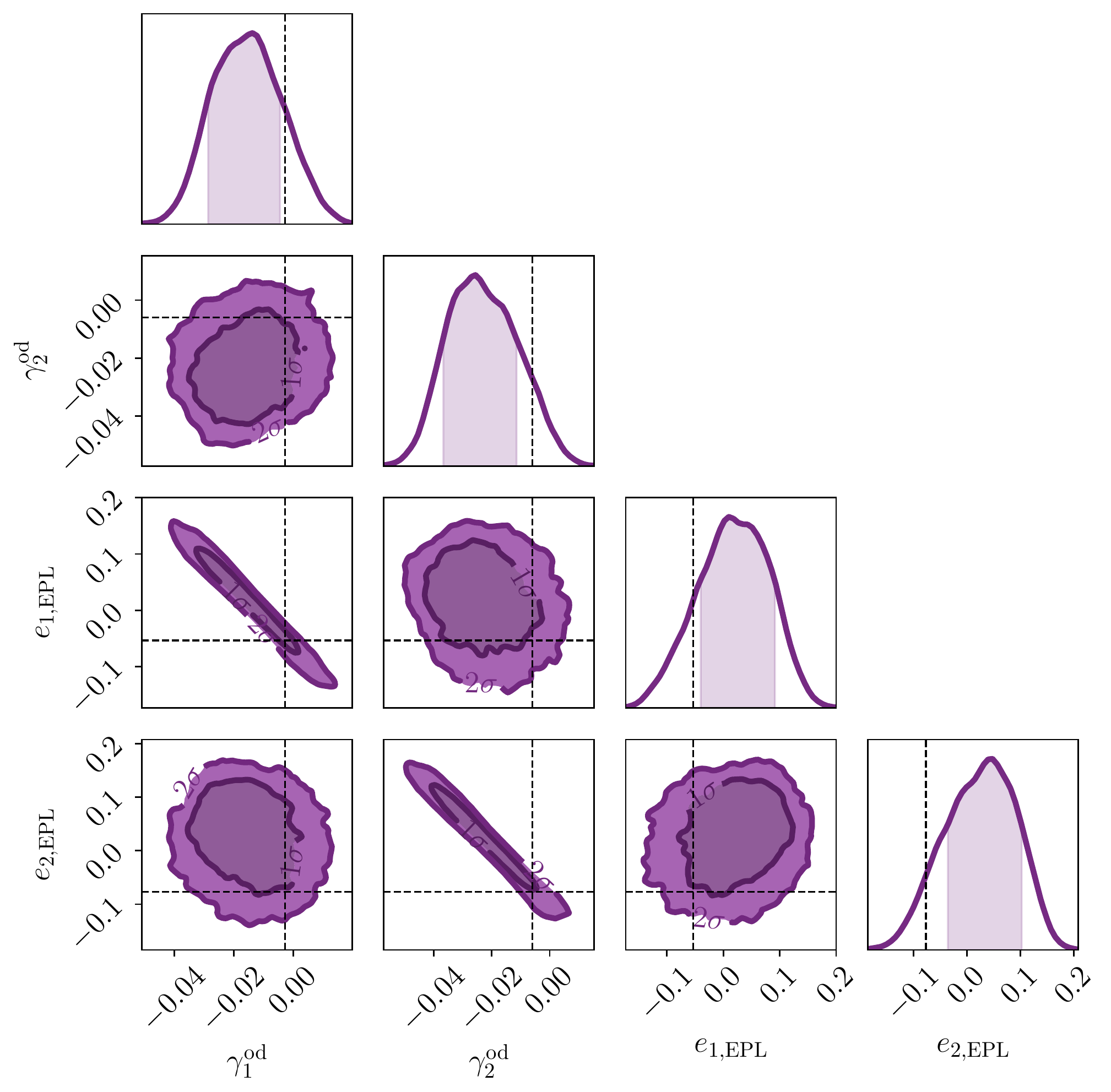}
    \caption{\textbf{Inference with the full LOS model~\eqref{eq:lens_eqn_los}}, from fitting the simulated image in \cref{fig:advantage_image}. \emph{Left panel}: one- and two-dimensional marginalised posterior distributions of the (os) and (ds) shears. \emph{Right panel}: Same for the (od) shear and lens ellipticity. The dashed lines show the input value of each parameter.}
    \label{fig:advantage_contours_full}
\end{figure*}

\begin{figure*}
    \centering
    \includegraphics[width=0.49\textwidth]{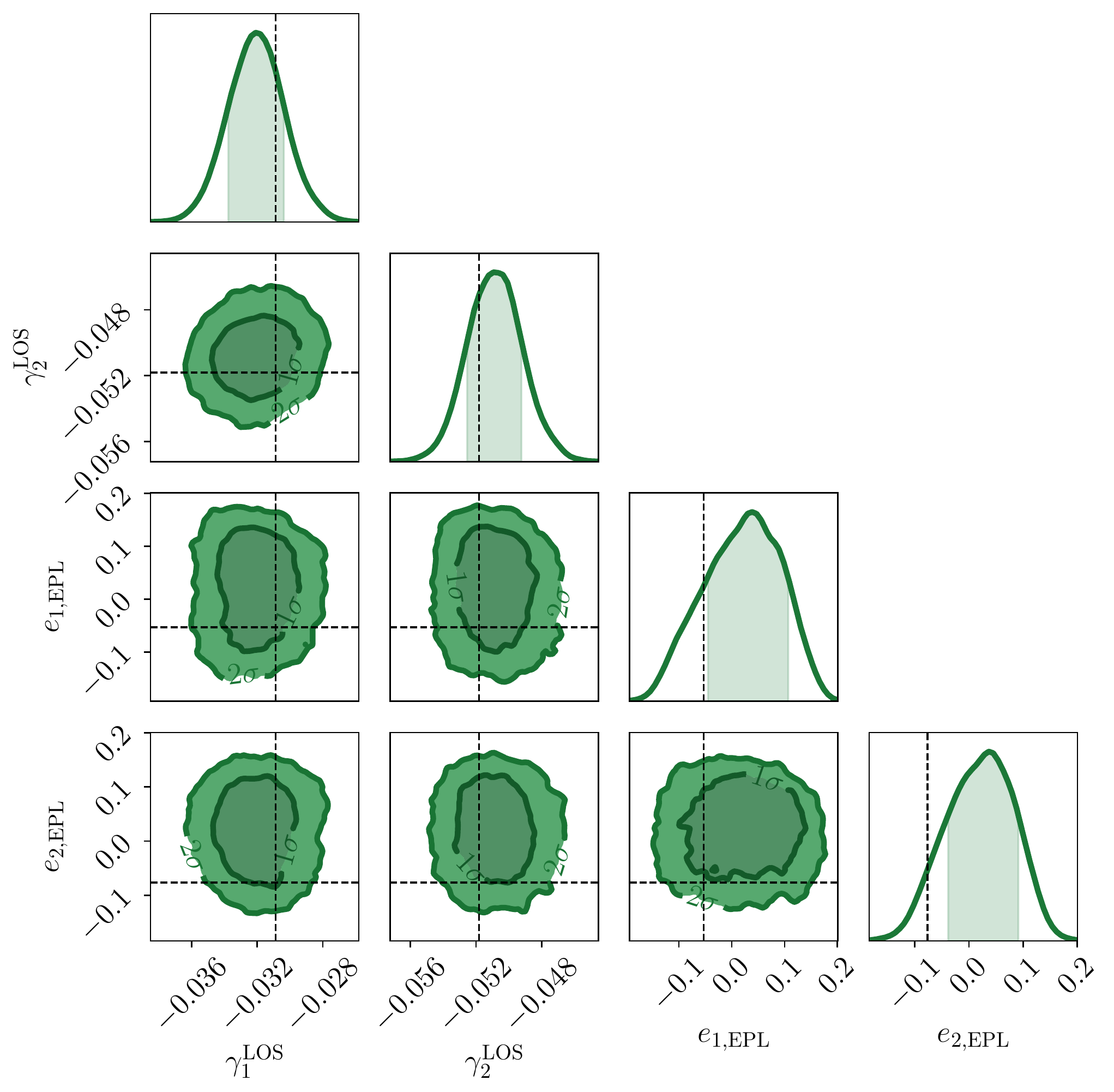}
    \hfill
    \includegraphics[width=0.49\textwidth]{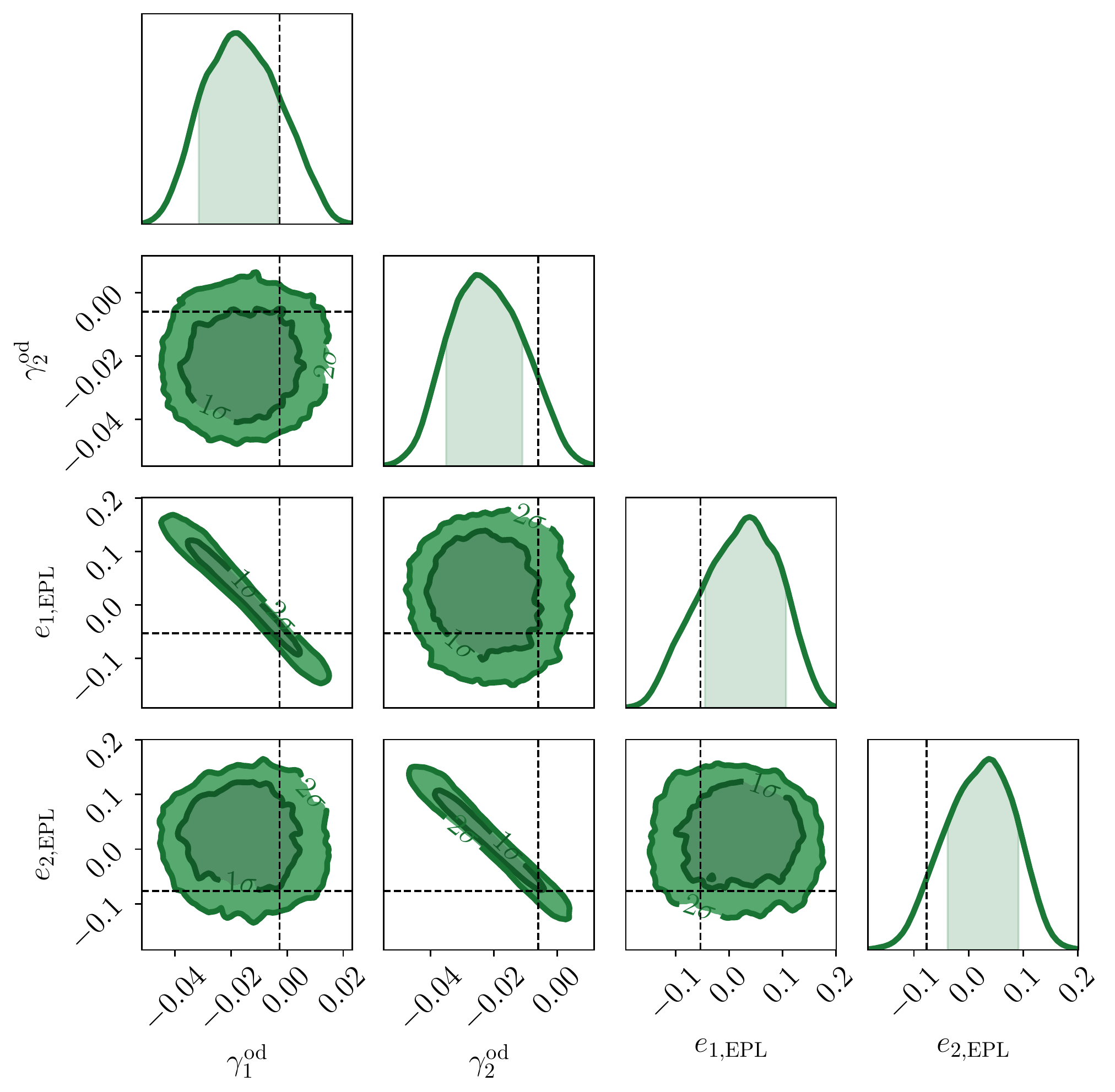}
    \caption{\textbf{Inference with the \minmodel~\eqref{eq:lens_eqn_minimal}}, from fitting the simulated image in \cref{fig:advantage_image} with the \minmodel. \emph{Left panel}: one- and two-dimensional marginalised posterior distributions of the effective LOS shear and lens ellipticity. \emph{Right panel}: Same for the (od) shear and lens ellipticity. The dashed lines show the input value of each parameter.}
    \label{fig:advantage_contours_minimal}
\end{figure*}

In \cref{fig:advantage_contours_full}, we show the one- and two-dimensional marginalised posterior distributions for the (os) and (ds) shears (left panel) and (od) shear and lens ellipticity (right panel) which result from fitting the simple simulated image with the full LOS model. Note that all the contour plots presented in this paper were produced using the \texttt{chainconsumer} package\footnote{\url{https://github.com/Samreay/ChainConsumer}.} \citep{Hinton2016}. 

From these plots, we can see that while the shear parameters are correctly recovered at the level of $1\sigma$, there are very strong degeneracies between the (os) and (ds) shears. This results in large uncertainties for $\gamma\e{os}$ and $\gamma\e{ds}$, on the order of \modification{$10\,\%$}, i.e. comparable or larger than the expected signal. Furthermore, the expected degeneracy between the (od) shear and the lens ellipticity is clearly present. As discussed in \cref{subsec:minimal_model}, the simplicity of the EPL model artificially alleviates that degeneracy, giving the impression that the main lens's ellipticity can be measured independently of the foreground shear.

In \cref{fig:advantage_contours_minimal}, we show the one- and two-dimensional marginalised posterior distributions for the (od) shear and lens ellipticity (left panel) and LOS shear and lens ellipticity (right panel) which result from fitting the simple simulated image with the \minmodel. From these plots, we can see that, as expected, the minimal LOS shear $\gamma\e{LOS}$ evades any of the previous degeneracies. The input value is perfectly recovered with, in this simple example, a remarkable precision on the order of a few parts in $10^{4}$.

The results presented in this section demonstrate the clear advantage of using the \minmodel over the full model. This advantage can be understood as being due to a more efficient parameterisation of the problem at hand. It is comparable to the use of $S_8\propto\sigma_8\sqrt{\Omega\e{m}}$ in standard cosmic shear, which is better constrained than $\sigma_8$ (see \citealt{2021MNRAS.505.4935H} for a thorough explanation as to why); or to the use of $\omega = \Omega h^2$ in cosmic microwave background constraints.

\section{Line-of-sight shear measurements from realistic mock images}
\label{sec:measurability_shear}

In this section, we show how the LOS shear is systematically measurable from a larger set of more complicated images than the simple example shown in the previous section. We also assess how the measurement of the LOS shear is affected if the images are fit with lens models that are slightly to drastically different to those used to create the images.

\subsection{Methodology}

We generate a catalogue of 64 mock images, depicted in \cref{fig:golden_sample_image}, following a protocol which we will describe in \cref{subsubsec:mock_generation}. We then fit those images with four different models which we present in \cref{subsubsec:fitting_models}, focusing on the inference of the shear~$\gamma\e{LOS}$. The precision and accuracy of the recovery $\gamma\e{LOS}$ for each model is assessed using statistical tools we will describe in \cref{subsubsec:tools}.

\subsubsection{Complex mock images from a composite lens model}
\label{subsubsec:mock_generation}

\begin{figure}
\centering
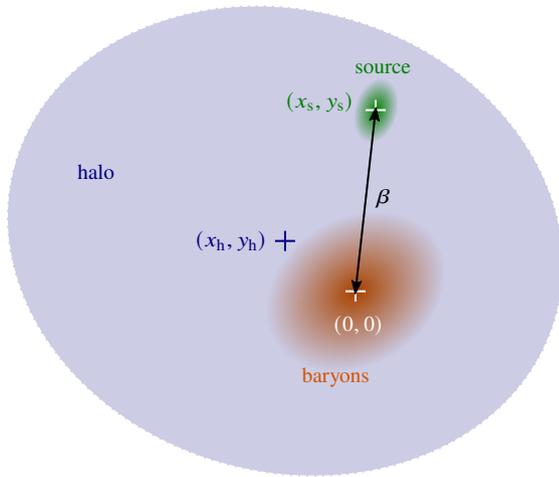
\caption{Schematic of our composite lens model on the celestial sphere. The baryonic component (orange) is described by a an elliptical Sérsic model whose centre is placed at $(0,0)$. The dark-matter halo (blue) has an elliptical NFW profile and its centre~$(x\e{h}, y\e{h})$ is offset with respect to the baryons. This composite main lens is then perturbed by the LOS shears $\gamma\e{od}, \gamma\e{os}, \gamma\e{ds}$. The source (green) is a perturbed elliptical Sérsic placed at $(x\e{s}, y\e{s})$.}
\label{fig:schematic_composite_lens}
\end{figure}

We increase the complexity of the lens modelling in a number of ways compared to \cref{sec:advantage}, so as to mimic real strong lensing images. We now model the lens with two constituent parts: a baryonic component modelled with an elliptical \Sersic profile, and a dark matter halo modelled with an elliptical Navarro--Frenk--White (NFW) profile \citep{Navarro:1995iw}. \modification{Note that in both profiles, the ellipticity is implemented at the level of the \emph{potential} rather than at the level of the convergence. This means that their iso-potential contours are elliptical, but not their iso-density contours.} The halo is offset from the centre of the baryons and the ellipticities of each component are not constrained to be of the same magnitude or direction. The set-up is depicted in \cref{fig:schematic_composite_lens}.

We model the lens and source light with elliptical \Sersic profiles. Furthermore, we perturb the source light by adding three additional elliptical \Sersic profiles on top of the original, each with an amplitude of \modification{1}\,\% of the main source. Such perturbations will not be included in the fitting model, and thus aim to account for our general ignorance of the details of the source light. Finally, we shear each image using the full LOS shear model. 

\begin{figure*}
    \centering
    \includegraphics[width=\textwidth]{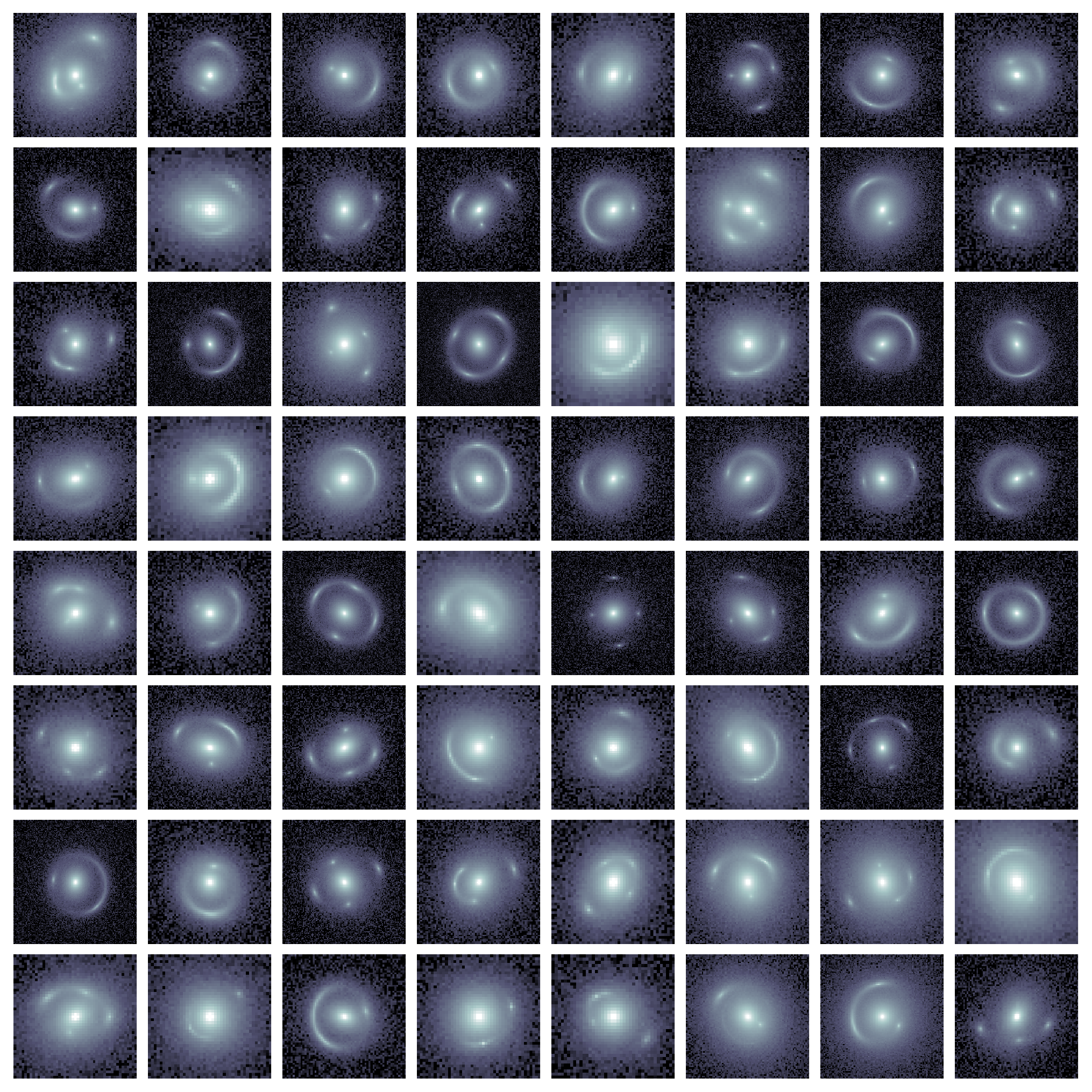}
    \caption{Our catalogue of 64 simulated images produced according to the protocol presented in \cref{subsubsec:mock_generation}.}
    \label{fig:golden_sample_image}
\end{figure*}

We simulate 64 such systems, each with every model parameter drawn at random from physically motivated distributions. The full parameter ranges and distributions are described in detail in \cref{appendix:composite_lens_parameters}. The resulting simulated images are shown in \cref{fig:golden_sample_image}, where their broad diversity in morphology can be clearly seen.

\modification{We quantify the quality of an image with a notion of cumulated signal-to-noise ratio. Specifically, for an image~$i$ composed of pixels~$p$, we define
\begin{equation}
Q_i \equiv \sum_{p\in i} \text{SNR}(p) \, \Theta[\text{SNR}(p) - 1] \ ,
\label{eq:quality_image}
\end{equation}
where $\text{SNR}(p)$ denotes the signal-to-noise ratio of pixel $p$. The lens light is not included in the signal because it is unrelated to the quality of the lensed image, and the noise accounts for both the background and shot noise. The Heaviside function $\Theta$ in the sum of \cref{eq:quality_image} implies that only the pixels with a signal-to-noise ratio larger than one are accounted for in the computation of the quality~$Q_i$. The advantage of this definition of quality is that it accounts for both the traditional notion of SNR (the brighter, the better) and for the number of pixels that are covered by the image (the larger and more resolved, the better).
}

\subsubsection{Four fitting models}
\label{subsubsec:fitting_models}

We employ four models to fit the simulated images in our catalogue, each with a different level of complexity compared to the mocks. We describe them below; see \cref{tab:summary_four_models} for a summary of the parameters included in or excluded from each model. 
\begin{enumerate}
\item The \emph{comprehensive model} is identical to the set-up used to generate the mock images, except that LOS effects are described by the \minmodel~\eqref{eq:lens_eqn_minimal} with parameters $\gamma\e{od}, \gamma\e{LOS}, \omega\e{LOS}$; the rotation term $\omega\e{LOS}$ is a nuisance parameter included to absorb the small non-linear effects arising from shear--shear couplings. As mentioned earlier, perturbations to the source light are not included.
\item The \emph{no foreground shear model} is identical to the comprehensive model, except that the foreground shear~$\gamma\e{od}$ involved in the effective potential of \cref{eq:effective_potential} is set to zero, thereby implying $\psi\e{eff}=\psi$ here. The motivation for this model is twofold. On the one hand, the presence of $\gamma\e{od}$ in the \minmodel is precisely what makes it different from a more traditional model comprising a composite lens plus external shear, such as those used for some of the H0LiCOW lenses~\citep{Wong:2019kwg}. It is thus essential to evaluate the role of this new parameter. On the other hand, we expect a degeneracy to exist between the ellipticity of the main lens and the foreground shear. If this degeneracy were perfect, then only two of the three complex parameters $\gamma\e{od}, e\e{\mSersic}, e\e{NFW}$ would suffice to describe an image, allowing us to set one of them to zero. The performance of the no foreground shear model will thus assess the validity of this degeneracy in a composite lens scenario.
\item In the \emph{aligned halo model}, the centres of the baryon and dark matter components are both fixed to $(0,0)$, thus removing the possibility for the halo to be offset from the baryons. The purpose of this model is to check if neglecting this offset leads to systematic biases in the inference of the other parameters, as was noticed by \cite{Gomer:2021gio}. The aligned model is otherwise identical to the comprehensive one.
\item In the \emph{power law model}, finally, the entire main lens is modelled using a single EPL profile, and with no additional NFW halo, as we used in our simple example in \cref{sec:advantage}. This model aims to quantify the impact of neglecting the composite character of a lens on the recovery of the LOS shear.
\end{enumerate}

\begin{table*}
\centering
\caption{Summary of the properties and results of the four models used to fit the mock images of \cref{fig:golden_sample_image}.}
\begin{tabular}{SlScScScSc}
\hline
\hline
& Comprehensive model & No foreground shear model & Aligned model & Power law model \\
\hline
Features & & & & \\
\hdashline[0.5pt/1pt]
Foreground shear    & \checkmark    & \xmark        & \checkmark    & \checkmark \\
LOS shear           & \checkmark    & \checkmark    & \checkmark    & \checkmark \\
Baryon ellipticity  & \checkmark    & \checkmark    & \checkmark    & -- \\
Halo ellipticity    & \checkmark    & \checkmark    & \checkmark    & -- \\
Baryon--halo offset & \checkmark    & \checkmark    & \xmark        & -- \\
\hline
Results                          & & & & \\
\hdashline[0.5pt/1pt]
Productive images: $N_\productive$ & \modification{63} (\modification{98}\,\%)  & \modification{62} (\modification{97}\,\%)  & \modification{56} (\modification{88}\,\%)  & \modification{28} (\modification{44}\,\%) \\
Precision: $\bar{\sigma}$ & \modification{0.011}       & \modification{0.011}       & \modification{0.010} & \modification{0.010}\\
Accuracy: $\chi^2$        & \modification{1.0}         & \modification{3.4}         & \modification{62.9}  & \modification{92.2} \\
Outliers: $N_{>2\sigma}$  & \modification{0}           & 2                 & \modification{36}    & \modification{20} \\
\hline
\end{tabular}
\label{tab:summary_four_models}
\end{table*}

\subsubsection{Evaluating the precision and accuracy of the models}
\label{subsubsec:tools}

We now present the quantitative criteria that we use to assess the precision and accuracy of the inference of $\gamma\e{LOS}$ in each model. Firstly, we note that when fitting the images in our mock catalogue, the MCMC parameter inference does not always succeed in obtaining either an upper or lower bound on one or both components of $\gamma\e{LOS}$, particularly in the less informative models. We therefore discard the cases where one or both bounds were not obtained from the quantitative comparison of the models that will follow. We refer to the remaining images in the mock catalogue as the \textit{productive images}, $\productive$, denoting their number as $N_{\productive}$.

Furthermore, for an image~$i\in\productive$ with a given model, the posterior distribution of $\gamma\e{LOS}$ is generally non-Gaussian; in particular, the $68\%$-confidence domain is not always symmetric, which translates into asymmetric error bars. We call $\sigma_{a,i}^\pm$ the corresponding upper and lower uncertainties at $68\%$-confidence level on the component~$a\in\{1,2\}$ of $\gamma\e{LOS}$ for the $i\h{th}$ image. \modification{This implies that for each image and model, the uncertainty on the LOS shear is a priori described by four numbers, which is somewhat cumbersome. For further simplicity, we summarise them all by the geometric mean of the symmetrised uncertainty of each component of the LOS shear:
\begin{equation}
\label{eq:mean_uncertainty}
\sigma_i
\equiv
\sqrt{
    \left( \frac{\sigma_{1,i}^+ + \sigma_{1,i}^-}{2} \right)
    \left( \frac{\sigma_{2,i}^+ + \sigma_{2,i}^-}{2} \right)
    } \ .
\end{equation}
The general \emph{precision} of a model is then quantified from the mean uncertainty on the LOS shear over all its productive images,
\begin{equation}
\label{eq:uncertainty_i}
\bar{\sigma} \equiv \frac{1}{N_\productive} \sum_{i\in\productive} \sigma_i \ .
\end{equation}
}

The \emph{accuracy} of a model, i.e. its ability to recover the true value of $\gamma\e{LOS}$, may be evaluated with a quantity inspired by the reduced $\chi^2$, accounting for the asymmetry of the uncertainties. Namely, for each component $a\in\{1, 2\}$ of the LOS shear, we define
\begin{align}
\chi^2_a
&\equiv
\frac{1}{N_\productive}
\sum_{i\in\productive}
\frac{1}{\sigma_{a,i}^2}
\left(\gamma\h{LOS, out}_{a,i} - \gamma\h{LOS, in}_{a,i}\right)^2
\label{eq:chi2}
\intertext{with} 
\sigma_{a,i}
&\equiv
\begin{cases}
\sigma_{a,i}^+ & \text{if}\ \gamma\h{LOS, out}_{a,i} \leq \gamma\h{LOS, in}_{a,i} \\
\sigma_{a,i}^- & \text{if}\ \gamma\h{LOS, out}_{a,i} > \gamma\h{LOS, in}_{a,i}
\end{cases}
\label{eq:sigma}
\end{align}
where $\gamma\h{LOS, in}_{a,i}$ and $\gamma\h{LOS, out}_{a,i}$ are respectively the true (input) value and the best fit (output) value of the component $a$ of the LOS shear of the $i\h{th}$ image. \modification{Again, for further simplicity and to avoid describing the accuracy of a model with two numbers $\chi^2_1, \chi^2_2$, we shall summarise them by their arithmetic mean,
\begin{equation}
\chi^2 \equiv \frac{1}{2} \left( \chi^2_1 + \chi^2_2 \right) .
\end{equation}
}

Lastly, we can examine the number~$N_{>n\sigma}$ of outliers, i.e. the images for which the best fit output for $\gamma\e{LOS}$ lies further than a certain number~$n$ of $\sigma$ from the true input value. Specifically, we consider that an image~$i$ is an $n\sigma$ outlier if
\begin{equation}
\left|\gamma\h{LOS, out}_i - \gamma\h{LOS, in}_i\right|^2
> n^2 \left(\sigma_{1,i}^2 + \sigma_{2,i}^2\right) \ , \label{eq:Nexcess}
\end{equation}
where $\sigma_{a,i}$ was defined in \cref{eq:sigma}. \modification{This definition thus accounts for the asymmetry of the error bars}.

\subsection{Results}

\subsubsection{Examining the fits of a single image}

\begin{figure*}
    \centering
    \includegraphics[width=0.49\textwidth]{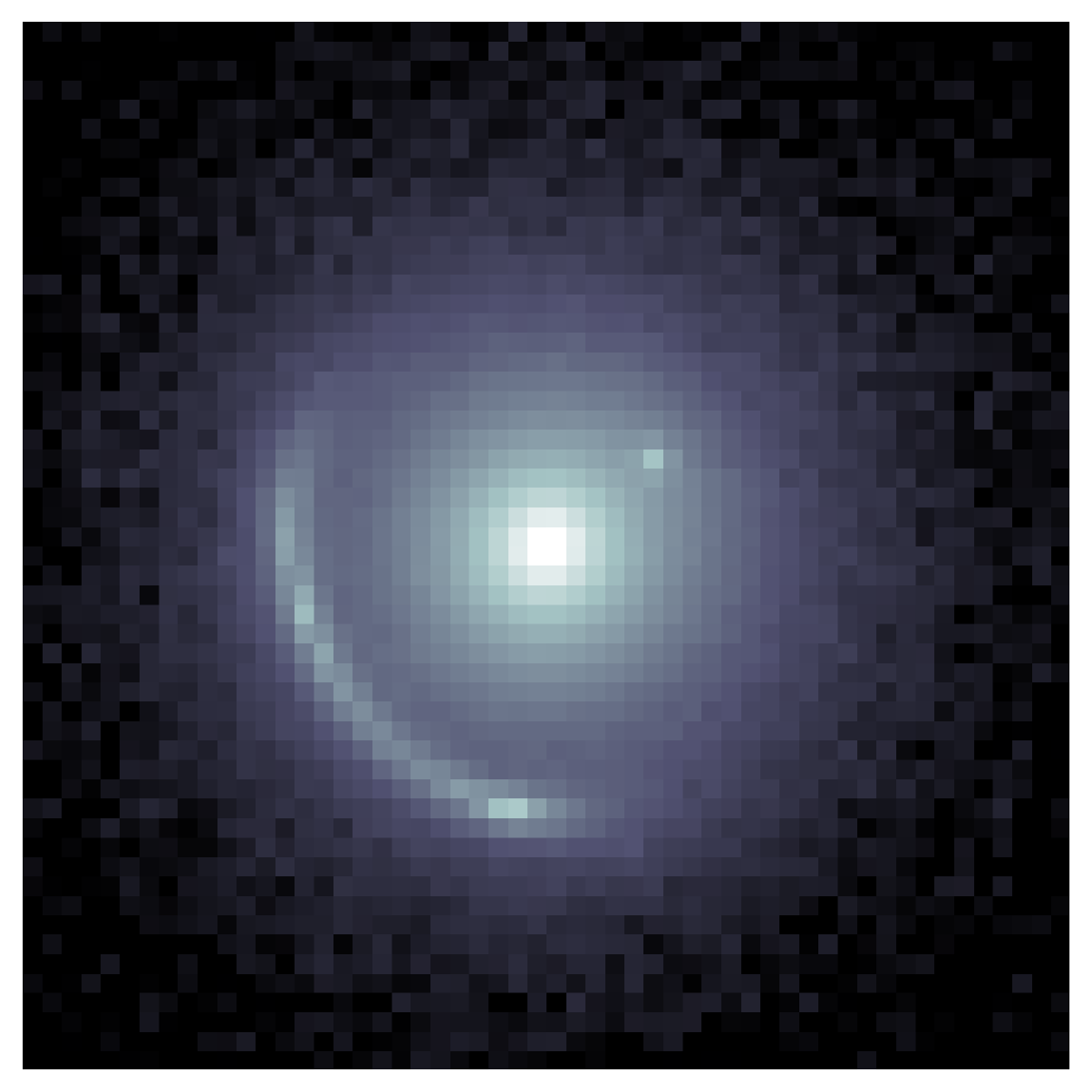}
    \hfill
    \includegraphics[width=0.49\textwidth]{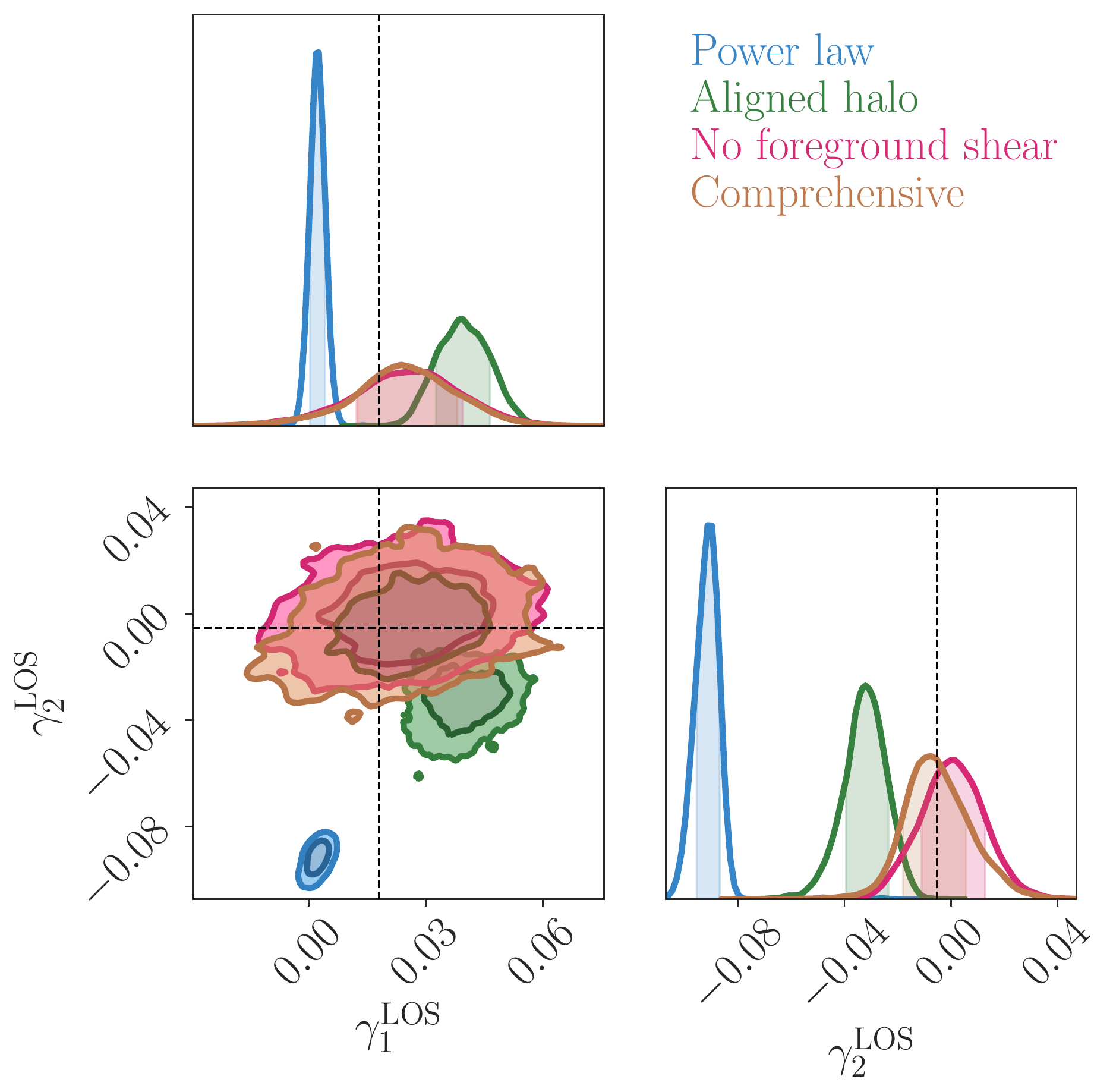}
    \caption{\textit{Left panel}: the \modification{\nth{44}} image in our catalogue. \textit{Right panel}: the one- and two-dimensional marginalised posterior distributions for $\gamma_1^{\rm LOS}$ and $\gamma_2^{\rm LOS}$ obtained by fitting this image with the power law model (blue), aligned halo model (green), no foreground shear model (magenta) and comprehensive model (orange). The dashed lines represent the expected values.}
    \label{fig:single_golden_result}
\end{figure*}

We firstly present the results of the fits to a single image as a representative example of the results in the rest of this section. In \cref{fig:single_golden_result} we show the \modification{\nth{44}} image in our mock catalogue, along with the one- and two-dimensional marginalised posterior distributions for $\gamma_1^{\rm LOS}$ and $\gamma_2^{\rm LOS}$ obtained when fitting with all four models described above. We chose this particular image to focus on as it is of roughly median quality and the shear parameters were well-sampled in all of the models used to fit it. 

From the contour plot, we can see that in both the comprehensive model and the no foreground shear model, the LOS shear is well recovered. On the other hand, in both the aligned model and in the power law model, the posteriors are fairly biased away from the expected values. This indicates that the less informative the model is, the worse the recovery of the shear is. However, the precision of the recovered shears is roughly the same in all four models, implying that the error bar on $\gamma\e{LOS}$ is insensitive to the model used. This is in line with the results we presented in \Cref{sec:advantage}, where we showed how $\gamma\e{LOS}$ is not degenerate with lens model parameters. We will explore this point further in the following.

\begin{figure*}
    \centering
    \includegraphics[height=0.91\textheight]{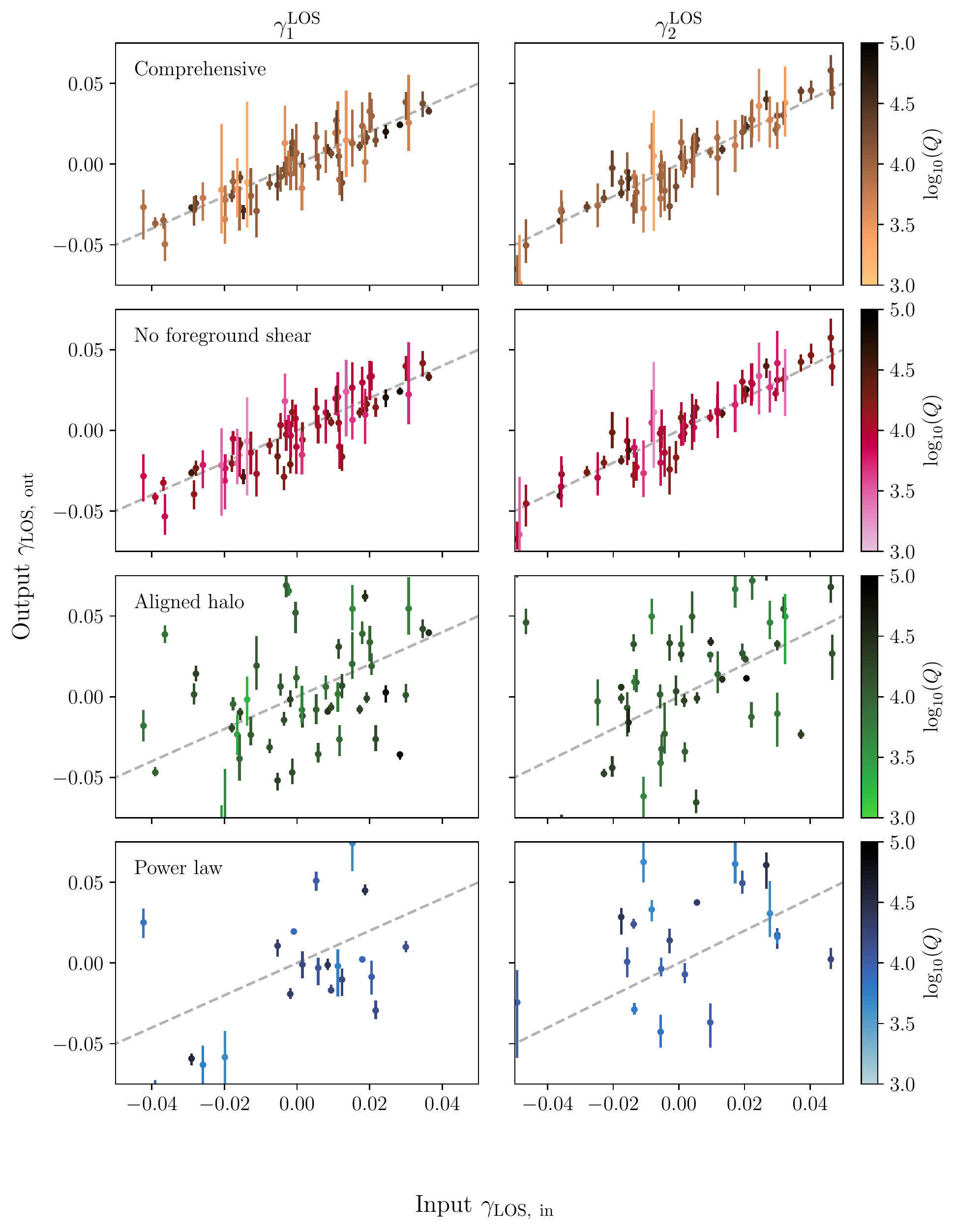}
    \caption{Input $\gamma\e{LOS, in}$ versus output $\gamma\e{LOS, out}$ in the comprehensive model (orange), no foreground shear model (magenta), aligned halo model (green) and power law model (blue). The left-hand column shows the results for $\gamma_1^{\rm LOS}$ and the right-hand column shows $\gamma_2^{\rm LOS}$. The dashed line in each panel is where $\gamma\e{LOS,~ in} = \gamma\e{LOS,~ out}$. The brightness of each point indicates the \modification{log of the quality criterion $Q$} for the associated system. A \modification{larger} $Q$ (\modification{darker} colour) corresponds to a \modification{larger SNR, or in other words, a better quality image.}}
    \label{fig:inout_all}
\end{figure*}


We now show the result of fitting our entire simulated image catalogue with the four models described above. In \cref{fig:inout_all}, we show the input LOS shear components as computed from the three individual shears of the full model versus the output LOS shear components as recovered from the images by the MCMC parameter inference. \modification{The colour bars show the log of the quality criterion $Q$. Darker colours thus represent larger values of $Q$ and hence larger SNR values.}

Only the results from the productive images in each model are shown in this figure.

\subsubsection{Results of the comprehensive model}

In the top two panels of \cref{fig:inout_all}, we show the result of fitting all the images in our catalogue with the comprehensive model. The number of productive images in this case is \modification{$N_{\productive}=63$, that is $98\,\%$} of the complete sample. We can see that in this model, the LOS shear is always very well recovered from the images.

The mean absolute precision of the recovered shear is \modification{$\bar{\sigma}= 0.01$} and does not show any correlation with the amplitude of $\gamma\e{LOS}$. In comparison, the precision of the inference of $\gamma\e{LOS}$ found in \cref{sec:advantage} when demonstrating the advantage of the \minmodel was about two orders of magnitude better. The better precision obtained in that case can be attributed to the relative simplicity of the lens and source model being used.

Our measure for the accuracy of the LOS shear recovery yields \modification{$\chi^2=1.0$}, which shows that uncertainties are well estimated for both components. \modification{As for the number of outliers, we find $N_{>2\sigma}=0$ points away from the true value by more than $2\sigma$. This is well within the expected proportion of $2\sigma$ outliers in the Gaussian case, $5\,\%$}. 

Furthermore, \modification{the image-quality criterion~$Q$ defined in \cref{eq:quality_image} is} well correlated with the size of the error bars on each point, with the most precise constraints generally coming from the best images (the darkest coloured points). 

\subsubsection{Results of the model with no foreground shear}

Next, we fit our catalogue of mock images with the model containing no foreground shear, $\gamma\e{od}=0$. The input--output plot for this case is shown in the second two panels of \cref{fig:inout_all}, with the points plotted in magenta. The number of productive images in this case is \modification{one fewer} than in the comprehensive model, $N_{\productive}=62$, that is $97\%$ of the catalogue.

With the exception of \modification{two outliers} (which are not visible due to the scale of the plot), the two LOS shear components are generally well recovered in this case too. The precision measure is unchanged compared to the comprehensive model, with \modification{$\bar{\sigma}=0.01$}. The accuracy is slightly \modification{reduced} overall, \modification{$\chi^2 = 3.4$}, \modification{while the number of outliers is increased}, with \modification{$N_{>2\sigma} = 2$ ($3\,\%$ of the productive images)}, \modification{though still} in line with the expectation for a normal distribution. This implies that the inclusion of foreground shear in the comprehensive model is not strictly necessary for a better fit to an image. We will discuss this point further below.

\subsubsection{Results of the aligned model}

Thirdly, we fit our catalogue of images with the aligned halo model. The input--output plot for this case is shown in the third two panels of \cref{fig:inout_all}, with the points plotted in green. We now see how worsening the model can degrade the fit; there is a larger number of outliers and unconstrained points, leading to the number of productive images shrinking to \modification{$N_{\productive} = 56$, that is $88\,\%$} of the catalogue. While the precision of the recovery of the two components of $\gamma\e{LOS}$ is \modification{unaffected, $\bar{\sigma}=0.01$}, the accuracy is now very bad, \modification{$\chi^2 = 62.9$}, with more than \modification{two thirds} of the productive images being outliers, $N_{\excess} = 36$.

\subsubsection{Results of the power law model}

Lastly, we fit our catalogue of images with the power law model. The input--output plot for this case is shown in the bottom two panels of \cref{fig:inout_all}, with the points plotted in blue. Of the four models used to fit the simulated images, this is clearly the worst, with a large number of unconstrained points that result from the difficulty the MCMC has when trying to fit complex images with an overly simplistic model: \modification{$N_{\productive}=28$}. Of the remaining points, the accuracy measure is \modification{$\chi^2=92.2$}, with \modification{again over two thirds} of the productive images leading to a measurement of $\gamma\e{LOS}$ that is biased by more than $2\sigma$, \modification{$N_{\excess} = 20$}. The mean precision of the recovered shear components \modification{remains in line with that of the other models,} $\bar{\sigma}=0.01$.

\subsection{Discussion}
\label{subsec:discussion}

\begin{figure*}
    \centering
    \includegraphics[width=\textwidth]{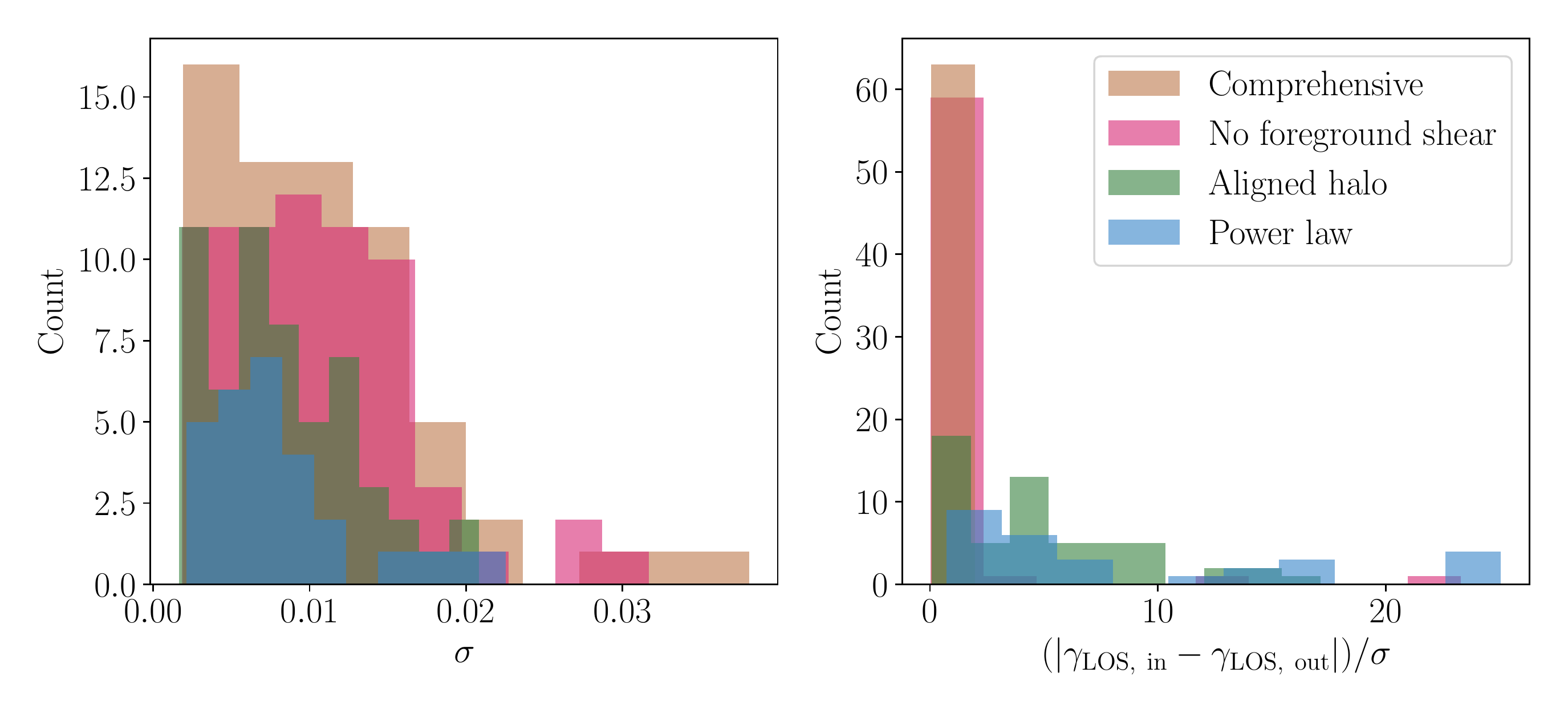}
    \caption{\textit{Left panel}: histogram of the uncertainty~$\sigma$ on $\gamma_{\rm LOS}$, as defined in \cref{eq:mean_uncertainty} for the four models. \textit{Right panel}: histogram of the difference between the input and best-fit output values of $\gamma_{\rm LOS}$ in units of $\sigma$.}
    \label{fig:error_histograms}
\end{figure*}

From these results, summarised in \cref{tab:summary_four_models}, we can clearly see the importance of using a sufficiently feature-rich model to fit complicated images, if a reliable and precise measurement of the LOS shear is desired, with well-sampled posterior distributions that yield upper and lower bounds on the recovered shear parameters. This is emphasised in \cref{fig:error_histograms}, where we plot the histogram of the mean error on the shear in each model (left panel) and the histogram of the difference between the input and output values of the shear (right panel). From this plot, we can see how the distribution of the size of the error is extremely similar across all the models. \modification{However, in the right-hand panel, we can see how the comprehensive model yields consistently more accurate shear measurements than the rest, with the greatest difference between input and output shear being less than $2\sigma$; this is in stark contrast to the aligned halo and power law models, where the greatest difference grows as large as $20\sigma$. The accuracy of the no foreground shear model is very close to that of the comprehensive model, barring the two outliers previously mentioned above.}

\begin{figure*}
    \centering
    \includegraphics[width=\textwidth]{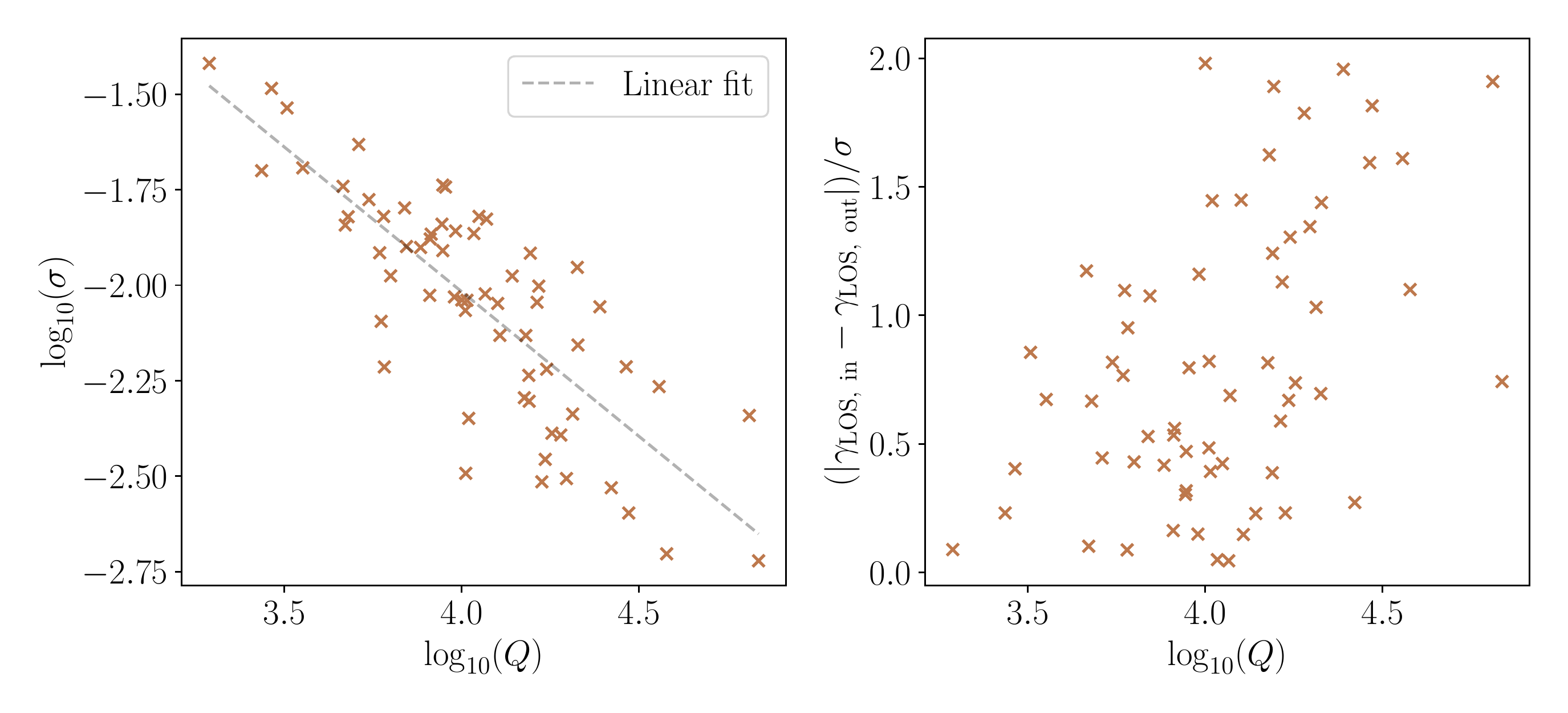}
    \caption{\textit{Left panel}: \modification{quality criterion $Q$ versus the estimated uncertainty $\sigma$ on $\gamma_{\rm LOS}$ [\cref{eq:mean_uncertainty}] in the comprehensive model. The grey dashed line represents a linear fit, \linfit,
    highlighting the correlation between these quantities}. \textit{Right panel}: \modification{quality criterion $Q$} versus the residuals of $\gamma_{\rm LOS}$ in units of $\sigma$ in the no foreground shear model. There is no significant correlation between these quantities.}
    \label{fig:error_vs_u}
\end{figure*}

The MCMC sampling is also less successful if a poor model is used: when using the no foreground shear model, we found two-sided bounds on both components the LOS shear for \modification{nearly $100\,\%$} of the catalogue, whereas when using the power law model, we found two-sided bounds on both the shear parameters for \modification{less than half} of the catalogue.

\modification{We can further examine the correlation between the quality measure $Q_i$ defined in \cref{eq:quality_image} and the uncertainty~$\sigma_i$ on $\gamma\e{LOS}$ defined in \cref{eq:mean_uncertainty} for each image $i$}, as plotted in the left-hand panel of \cref{fig:error_vs_u}. From this plot, we can see a strong \modification{negative} correlation between these quantities, with larger average errors coming from images of worse quality \modification{(smaller $Q$)}. A simple linear fit yields \modification{\linfit}. This implies that for precise measurements of $\gamma\e{LOS}$ to be obtained, poor quality images should be removed from the catalogue. On the other hand, in the right-hand panel of \cref{fig:error_vs_u}, we plot the difference between the input and output $\gamma\e{LOS}$ in units of $\sigma$ against $Q$. We can see that there is virtually no correlation between these quantities; in other words, the \emph{accuracy} of the shear recovery is not affected by the image quality. Note that we chose to plot these relationships for the \modification{comprehensive} model as an example, \modification{but the trend remains the same for the no foreground shear model}.

The excellent recovery of $\gamma\e{LOS}$ obtained from fitting with the no foreground shear model show that the degeneracy between the foreground shear and the properties of the main lens is indeed strong enough that the composite lens images can be satisfactorily obtained by the MCMC parameter inference even when $\gamma\e{od}$ is fixed to zero. This is another powerful argument in favour of the use of the \minmodel and again strongly emphasises the fact that $\gamma\e{LOS}$ is the only notion of shear that can evade such degeneracies and thus be accurately measured from strong lensing images.

\modification{For test purposes, we tried to increase the amplitude of the source perturbations, which we recall are not accounted for in the fitting models, so as to mimic a poorer modelling of the source light. In that case, we found that the reduced $\chi^2$, which quantifies the accuracy of the measurement of $\gamma\e{LOS}$, increases and becomes larger than unity even for the comprehensive model. This suggests that a too simple modelling of the lens light could lead to systematic biases in the measurements of the LOS shear. Such a systematic effect could be mitigated by using complex source models when fitting the images, but this approach may not translate easily when it comes to analysing real data \citep{Nightingale2015, Nightingale:2017cdh}.}

The results presented in this section show that the LOS shear is systematically measurable from simulated images if the lens model used to fit the images is of comparable complexity to that used to generate them. They also show that the foreground shear does not need to be included for good inference if the model for the main lens is sufficiently rich.

\section{Validity of the tidal approximation}
\label{sec:validity_tidal_regime}

The analysis and results presented so far were based on the lens equation~\eqref{eq:lens_eqn_los}, which supposes that all LOS perturbations to a strong lens -- typically dark matter haloes that would lie near the optical axis -- can be modelled as tidal fields, whose effect is encoded in the amplification matrices $\amplification\e{od}, \amplification\e{os}, \amplification\e{ds}$. In this section, we assess the validity of this tidal approximation by fully simulating the effect of randomly distributed haloes on an image.

\subsection{The tidal regime and beyond}

A LOS perturber is said to be in the tidal regime if the deflection field that it produces can be described by a quadratic potential across the strong lensing image under consideration. Suppose for instance that the observer sees a source that is lensed by a halo $h$ in the absence of the main lens and any other perturber. The corresponding Fermat potential can be expanded at second order in the position~$\bm{\theta}$ in the vicinity of the LOS ($\bm{\theta}=\bm{0}$),
\begin{equation}
\label{eq:potential_second_order}
\psi_h(\bm{\theta})
= \psi_h(\bm{0})
    + \bm{\alpha}_h(\bm{0}) \cdot \bm{\theta}
    + \frac{1}{2} \bm{\theta} \cdot \amplification_h \bm{\theta}
    + \ldots
\end{equation}
where $\bm{\alpha}_h(\bm{0})$ is the displacement angle caused by $h$ for an image observed along the optical axis, and $\amplification_h$ is the contribution of $h$ to the amplification matrix~$\amplification\e{os}$. A similar expansion could be performed for the (od) and (ds) cases; see \cite{Fleury:2021tke} for further details on the general theory of LOS perturbations in strong lensing. We have chosen to drop the (os) subscripts or superscripts when there is no ambiguity to alleviate notation.

The tidal regime for the halo $h$ is satisfied as long as \cref{eq:potential_second_order}, together with its (od) and (ds) analogues, remain valid up to an angle on the order of the main lens's Einstein radius, $\theta\e{E}$. This requirement is satisfied when $h$ lies sufficiently far from the LOS; as it gets closer, the higher-order terms hidden in the suspension points of \cref{eq:potential_second_order} may become non-negligible. Among them, the first one that we may worry about is the cubic term, $\mathcal{O}(\theta^3)$, generally referred to as \emph{flexion} because its effect is to produce arc-shaped images \citep{Bacon:2005qr}. Flexion can be understood as gradients of $\amplification$, i.e. gradients of convergence and shear; it consists of two complex numbers\footnote{These definitions follow the conventions of \cite{Fleury:2021tke}, which differ by a factor two from those of \cite{Bacon:2005qr}.}
\begin{equation}
\label{eq:flexion_def}
\mathcal{F} \equiv \partial \gamma = \partial^* \kappa \ ,
\qquad
\mathcal{G} \equiv \partial^* \gamma \ , 
\end{equation}
where $\partial\equiv(\partial/\partial\theta_1 - \ii \partial/\partial\theta_2)/2$ is the standard complex derivative with respect to the complex counterpart $\theta_1+\ii\theta_2$ of $\bm{\theta}=(\theta_1, \theta_2)$. When the flexion due to LOS perturbers is properly accounted for in the lens equation, \cref{eq:lens_eqn_los} is significantly more complicated, with the addition of four extra complex parameters~\citep{Fleury:2021tke}.

\subsection{A line of sight populated with haloes}

Since all the previous results we presented were obtained under the assumption that the tidal approximation is valid, we need to understand if it is a good description of the real Universe: how often do perturbers escape this regime, and do they jeopardise our ability to measure the LOS shear from strong lensing images?

In order to test this, we simulate a simple strong lens with a very large number of dark matter haloes in a volume around and along the LOS using the multi-plane lensing formalism. We compute the shear induced on the image by these LOS haloes and then fit the image using the same MCMC procedure followed in the previous sections, with a model again based on the \minmodel. If the tidal approximation is valid, we expect \modification{the model to provide a good fit and an accurate recovery of the induced shear from the image}. If the tidal approximation is significantly broken, we expect to see some sign of it, which could be a biased recovery of the shear, or a bad fit due to the non-negligible role of higher-order distortions.

The main lens and source light are modelled in the same way as in \cref{sec:advantage}, except that, for the sake of simplicity, we do not include any lens light. The ellipticities of the lens and source are drawn at random, while the remaining parameters are chosen in a physically realistic way. Specifically, we set the Einstein radius of the main lens \modification{$\theta_{\rm E}= 1.5''$}, the slope of the elliptical power law potential $\gamma_{\rm EPL}=2.6$, the apparent magnitude of the source as \modification{$23$}, the half-light radius of the source \modification{$R_{\rm \mSersic}=0.3''$ } and the \Sersic index \modification{$n_{\rm \mSersic}=6.0$}. The main lens is placed at $z_{\rm d}=0.5$ and the source is placed at $z_{\rm s}=1.5$. The angular positions of the centre of the main lens and of the source are chosen so as to be observed at $\bm{\theta}=\bm{0}$ in the presence of the haloes, $\bm{\beta}\e{s}=-\bm{\alpha}\e{os}(\bm{0})$ and $\bm{\beta}\e{d}=-\bm{\alpha}\e{od}(\bm{0})$. Conversions between distances and redshifts are performed with \texttt{astropy} \citep{astropy:2022}. As a reminder, we work in the spatially flat $\Lambda$CDM cosmology described by the \emph{Planck} 2018 data \citep{Aghanim:2018eyx}, with $H_0 = 67.4$ km s$^{-1}$ Mpc$^{-1}$ and $\Omega_{\rm m } = 0.315$.

\begin{figure}
    \centering
    \includegraphics[width=\columnwidth]{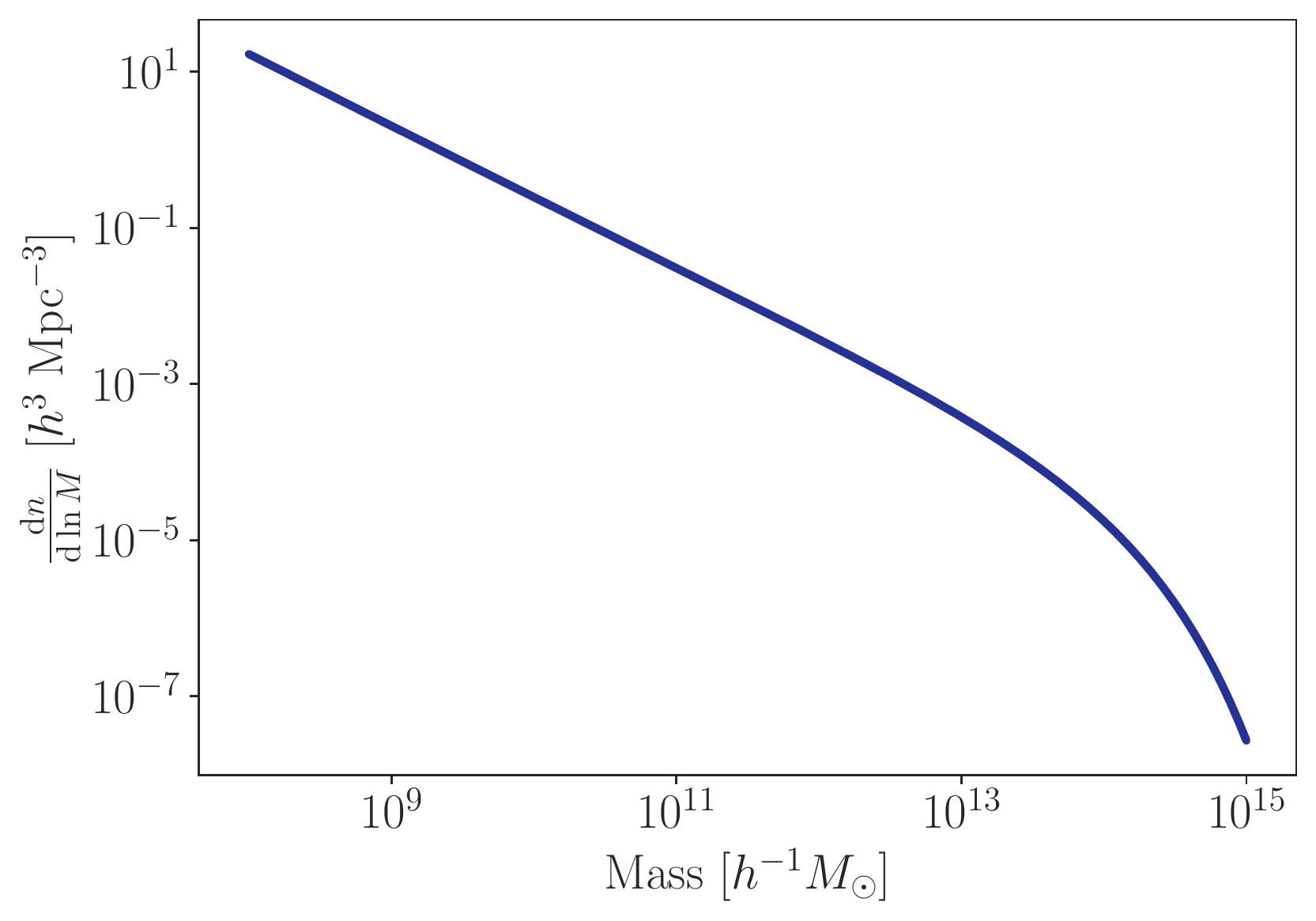}
    \caption{The \protect\cite{Tinker:2008ff} halo mass function between $10^{8}\,M_\odot$ and $10^{15}\,M_{\odot}$.}
    \label{fig:hmf}
\end{figure}

We generate the halo population as follows. We consider a region of space consisting of a comoving cylinder around the optical axis, with comoving radius $R=11.6\,\mathrm{Mpc}$. This number is chosen so that a point lens with mass $M= 10^{12}\,M_\odot$ located halfway between the observer and the source and with a transverse comoving distance $R$ from the optical axis would produce an (os) shear of $10^{-6}$ on the latter. We then randomly populate this region with a uniform comoving number density
\begin{equation}
n = \int_{M\e{min}}^{M\e{max}} \frac{\dd M}{M} \; \frac{\dd n}{\dd \ln M} \ ,
\end{equation}
where $\dd n/\dd\ln M$ denotes the halo mass function (HMF) evaluated following \cite{Tinker:2008ff} as implemented in the \texttt{colossus} package\footnote{\url{https://bitbucket.org/bdiemer/colossus.}} \citep{Diemer:2017bwl} at $z_{\rm d} = 0.5$. We fix a lower mass cutoff to $M\e{min}=10^8\,M_\odot$ corresponding to the typical mass of galactic subhaloes in the cold dark matter scenario, and $M\e{max}=10^{15}\,M_\odot$, which yields $n=5.6\,\mathrm{Mpc}^{-3}$. The mass of each halo is randomly drawn from the same HMF, shown in \cref{fig:hmf}.

\begin{figure}
    \centering
    \includegraphics[width=\columnwidth]{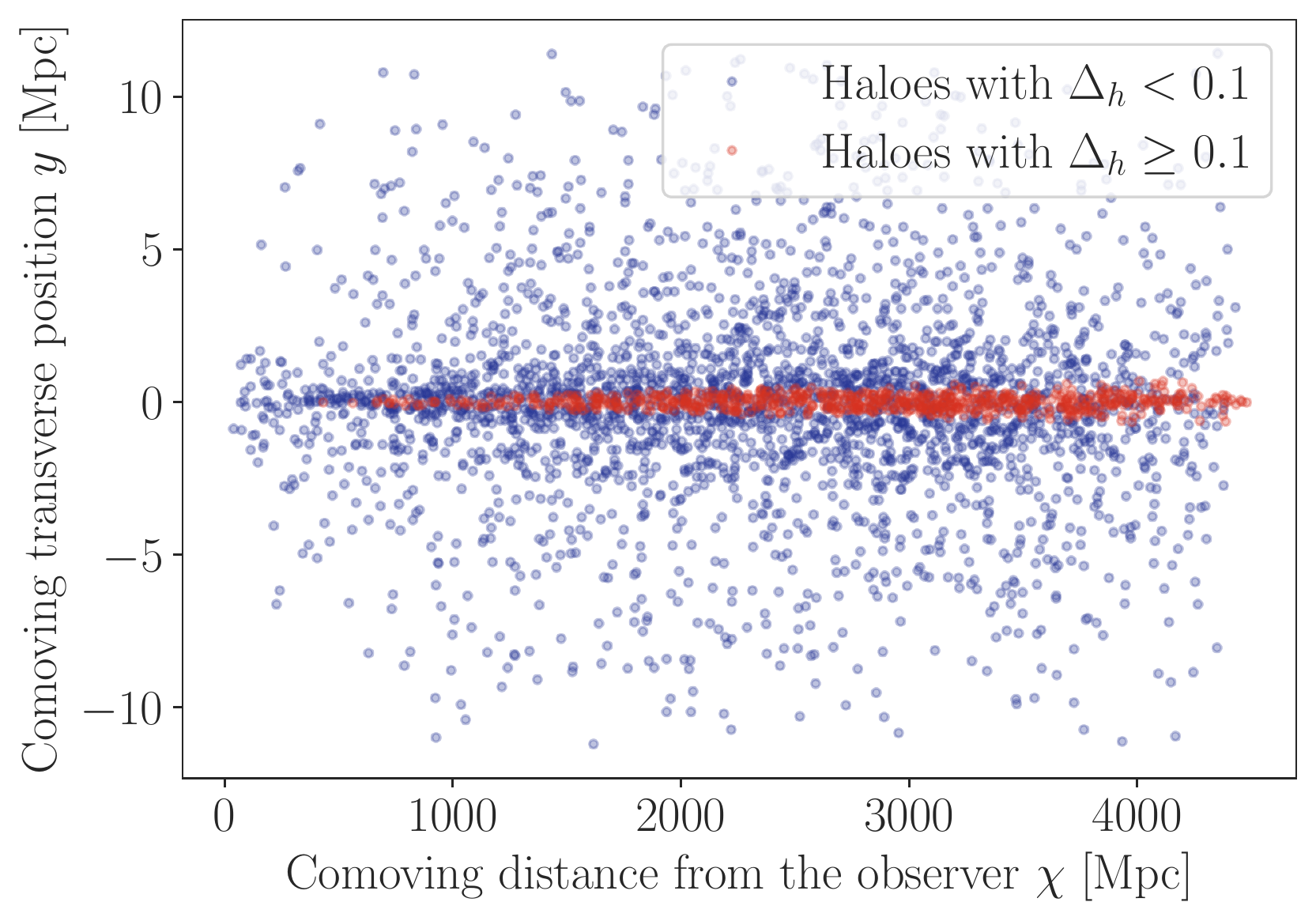}
    \caption{The cylindrical region populated with our sample of $N$ haloes, shown in projection in blue (for clarity only a tenth of the population is plotted), of which around 14\,\% have a relative variation of shear~$\Delta_h$ greater than 0.1 across $\theta\e{E}$ (of which we again plot a tenth), shown in red.}
    \label{fig:dist_haloes}
\end{figure}

This procedure results in a very large number of haloes, $N\e{tot}=1.1\times 10^7$ in the cylindrical region, among which many are practically too light or too far from the LOS to be relevant to the analysis. For computational efficiency, we thus remove the haloes for which a point lens with the same mass would individually produce an (os) shear with magnitude below $10^{-6}$ on the optical axis. After this operation, we are left with $N=3.4\times 10^4$ haloes , whose positions are depicted in \cref{fig:dist_haloes}. The points in red indicate the haloes that significantly evade the tidal regime (see \cref{subsec:validity_tidal_regime}).

\begin{figure*}
    \centering
    \includegraphics[width=0.32\textwidth]{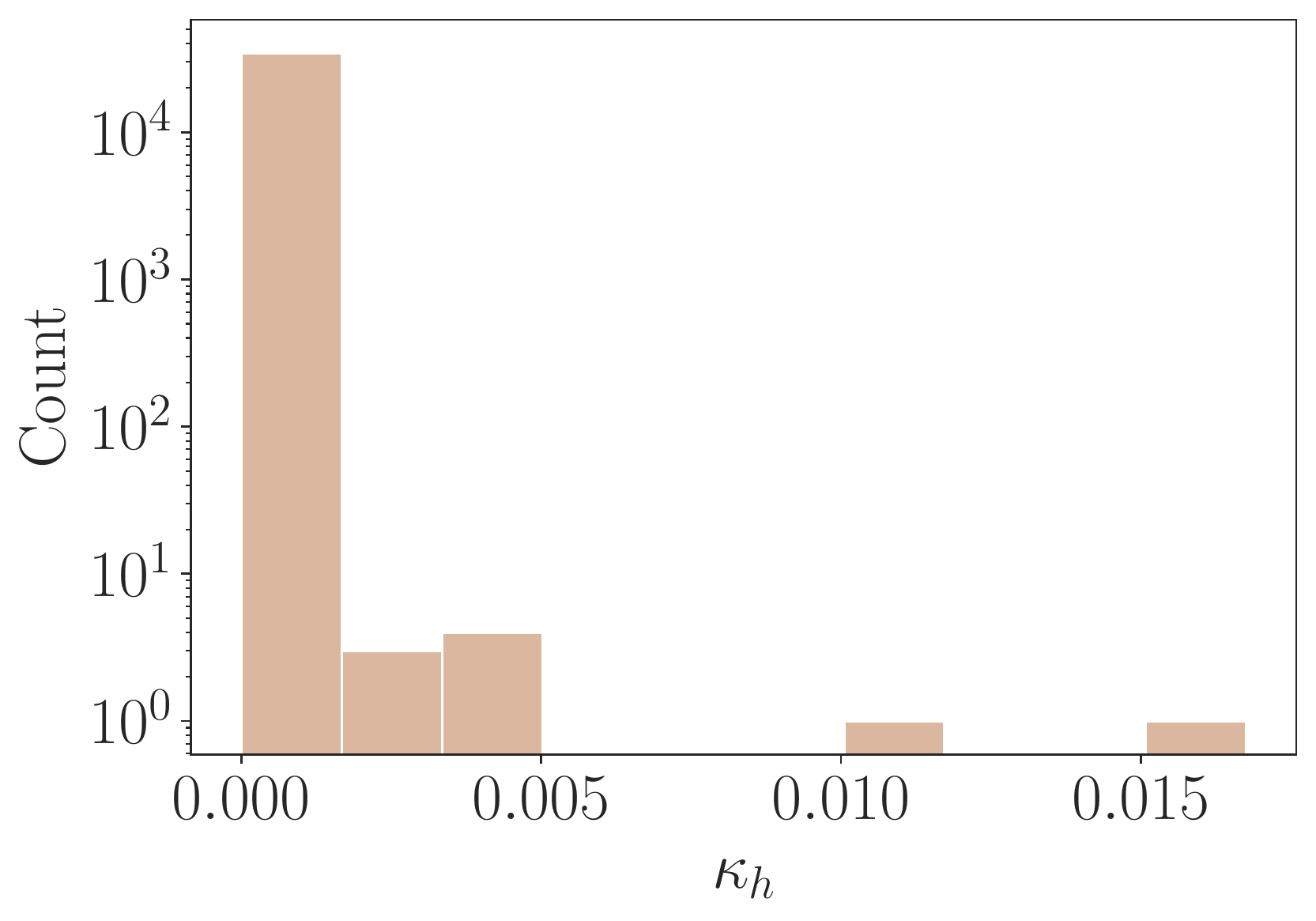}
    \includegraphics[width=0.32\textwidth]{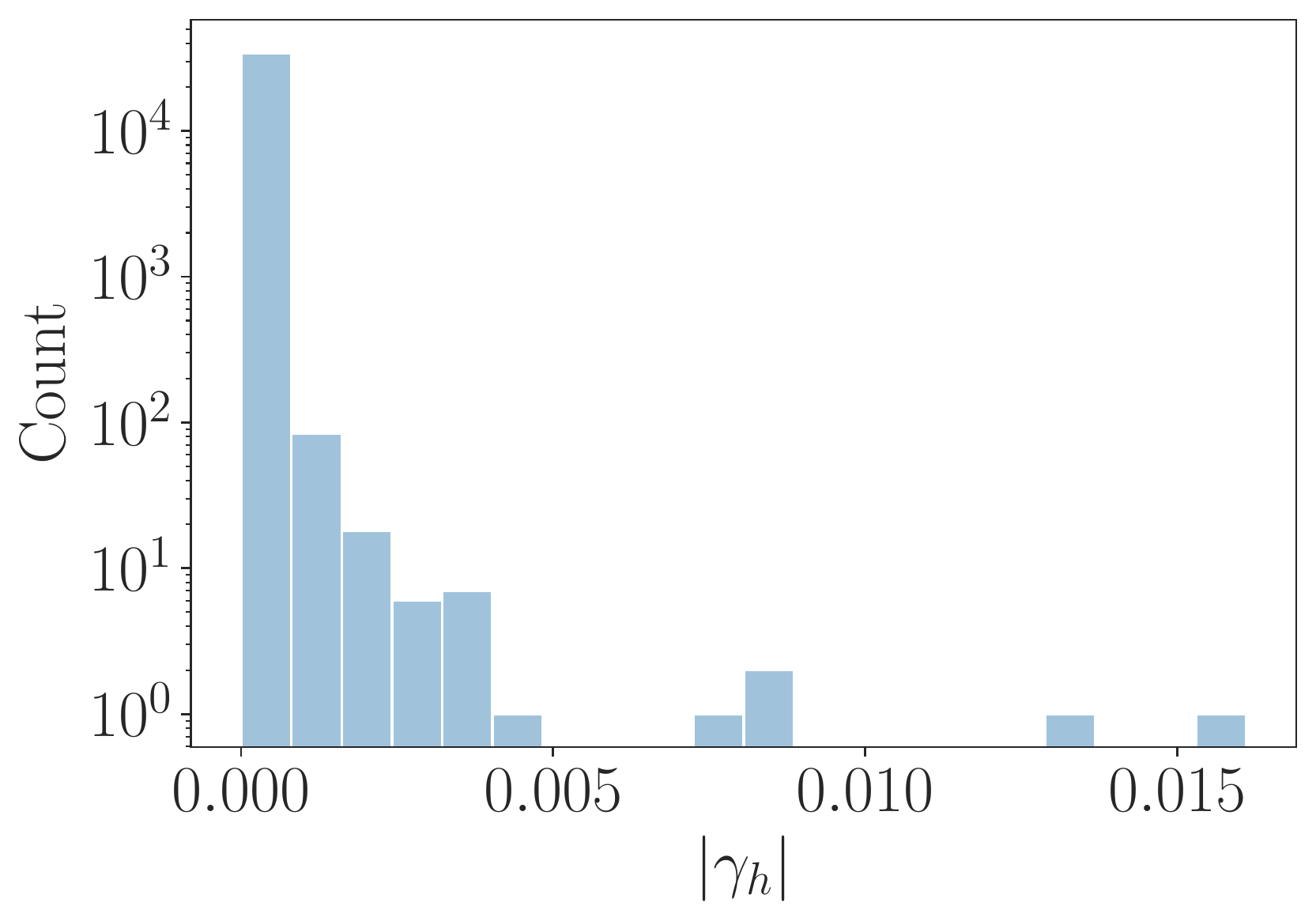}
    \includegraphics[width=0.32\textwidth]{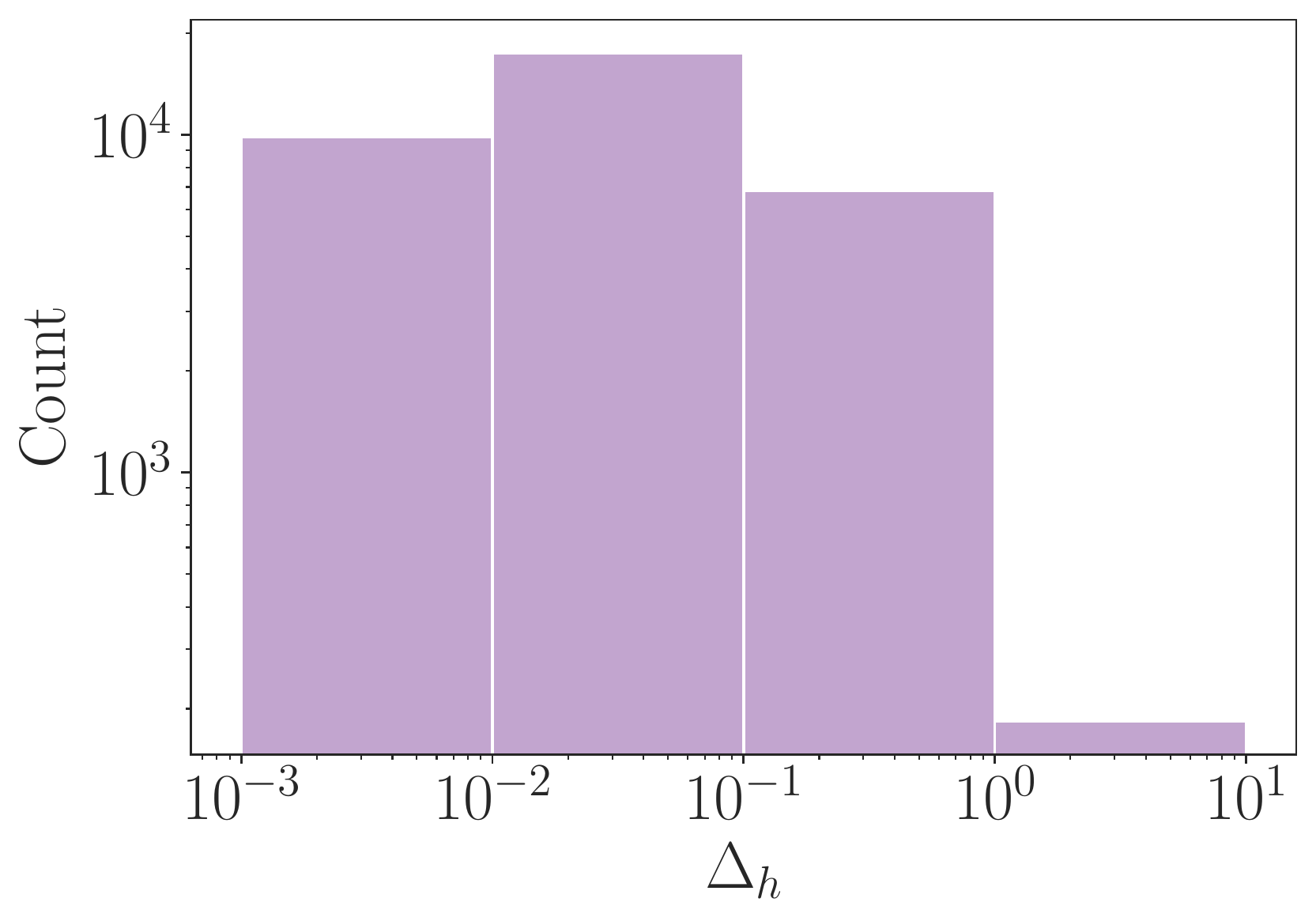}
    \caption{\textit{Left panel}: the histogram of the convergence $\kappa_h$ produced by individual haloes~$h$ on the optical axis. \textit{Middle panel}: same for the modulus of the shear $|\gamma_h|$. \textit{Right panel}: same for the relative variation of shear~$\Delta_h$ on the angular scale~$\theta\e{E}$. These are all the (os) quantities.}
    \label{fig:histograms}
\end{figure*}

The haloes themselves are modelled with NFW profiles. We compute their concentration using the \cite{Diemer:2018vmz} model for the mass--concentration relation, which is available in \texttt{colossus}. We then use the cosmology subpackage of \lens to convert these into the angular scale radius and the deflection angle at the scale radius which constitute the remaining parameters of the NFW mass profile in \lens. The left and middle panels of \Cref{fig:histograms} show the distribution of individual (os) shear and convergence produced by the $N$ haloes on the LOS. We can see that the individual tidal effects are generally small, except for a handful of haloes producing convergences and shears of percent order.

Lastly, in each plane containing a halo, we add a uniform component of negative density in order to compensate for the added mass under the form of haloes. This procedure mostly aims to avoid producing a too strong net convergence along the LOS. In practice, we set the negative convergence in each of those planes $h$ so as to exactly compensate the positive convergence~$\kappa_h(\bm{0})$ produced by the halo on the optical axis. This choice is not strictly consistent from a physical point of view, but it has the virtue of simplicity.

\subsection{Is the tidal regime a good approximation?}
\label{subsec:validity_tidal_regime}

\begin{figure*}
\centering
\includegraphics[width=\textwidth]{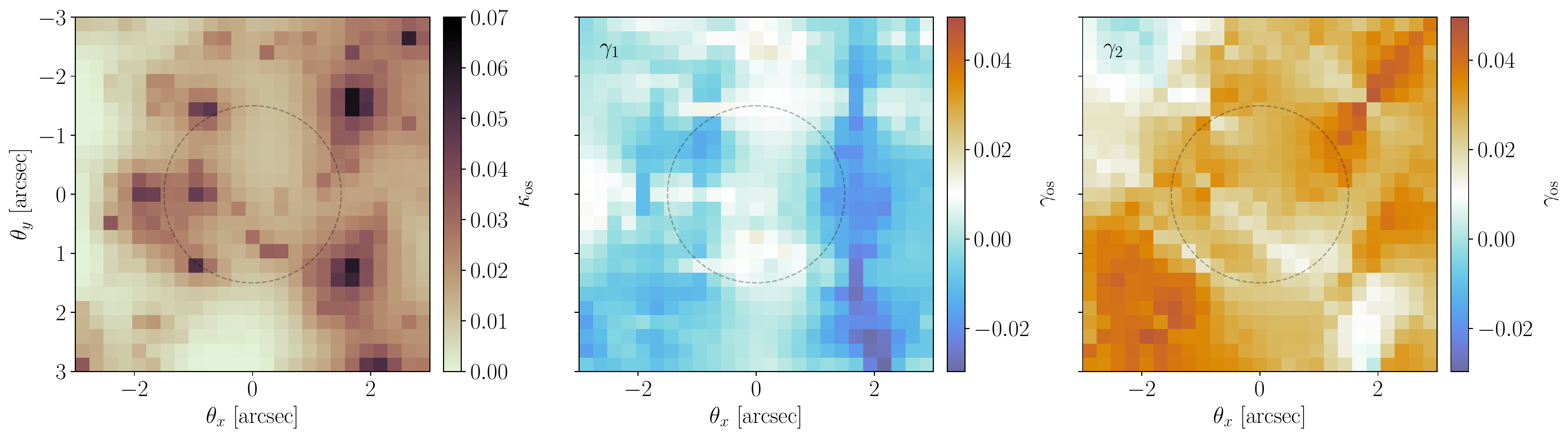}
\caption{\modification{Convergence field~$\kappa\e{os}(\vect{\theta})$ (\emph{left panel}) and shear field~$\gamma\e{os}(\vect{\theta})$ (\emph{middle and right panels}) produced by the randomly placed haloes around the LOS. Dashed circles indicate the Einstein radius of the main lens.}}
\label{fig:dist_haloes_shear_convergence_map}
\end{figure*}

\modification{A first, qualitative, assessment of the tidal approximation can be made by examining the variations of the convergence and shear fields due to the haloes only, i.e. in the absence of the main lens. This is shown in \cref{fig:dist_haloes_shear_convergence_map}, where we observe that the (os) convergence and shear due to the haloes display variations on the order of $100\,\%$ on the scale of the main lens's Einstein radius. The convergence map in particular (left panel of \cref{fig:dist_haloes_shear_convergence_map}) reveals the presence of a handful of haloes sitting very close to the optical axis, hence strongly evading the tidal regime, while remaining subcritical.

This diagnostic can be further quantified} by computing the relative variation~$\Delta$ of the (os) shear~$\gamma$ as we move away from the optical axis by one unit of the main lens's Einstein radius, $\theta\e{E}=1''$,
\begin{equation}
\Delta^2
= \left| \frac{\gamma(\theta_{\rm E}\,\hat{\bm{x}}) - \gamma(\bm{0})}{\gamma(\bm{0})} \right|^2
  + \left| \frac{\gamma(\theta_{\rm E}\,\hat{\bm{y}}) - \gamma(\bm{0})}{\gamma(\bm{0})} \right|^2 \ ,
\end{equation}
where $(\hat{\bm{x}}, \hat{\bm{y}})$ is an orthonormal basis of the image plane. Note that if the gradients of the shear can be considered constant across the image (flexion regime), then $\Delta$ can be expressed in terms of the two complex flexions defined in \cref{eq:flexion_def} as
\begin{equation}
\Delta^2 =
\frac12 \theta_{\rm E}^2 \, |\gamma|^{-2}
\left(|\mathcal{F}|^2 + |\mathcal{G}|^2 \right).
\end{equation}

We first consider the value~$\Delta_h$ of $\Delta$ for each halo $h$ taken individually. The distribution of $\Delta_h$ for the $N$ haloes is shown in the right panel of \cref{fig:histograms}. We can see that a considerable number of haloes produce a shear with significant variations on the scale of $\theta\e{E}$. \Cref{fig:dist_haloes} shows in red the haloes for which $\Delta_h \geq 0.1$; they are unsurprisingly clustered around the optical axis. We find that about $14\,\%$ of the total sample ($4.8\times 10^3$ haloes) have these especially large values of $\Delta_h$.

As expected from the above consideration, the combined effect of all the haloes does escape the tidal regime. We indeed find $\Delta\e{tot}=0.44$ for the total (os) shear, whose central value is $\gamma\e{os}=-2.4\times 10^{-2} + 3.8\ii\times 10^{-3}$. This number drops to $\Delta\e{tot}=0.045$ if we remove all the haloes~$h$ with $\Delta_h\geq 0.1$, in which case the cumulative shear becomes $\gamma\e{os}=-1.8\times 10^{-2} - 9.8\ii\times 10^{-4}$. This indicates that the haloes escaping the tidal regime are not negligible in the total shear produced along the LOS.

\begin{figure*}
    \centering
    \includegraphics[width=\textwidth]{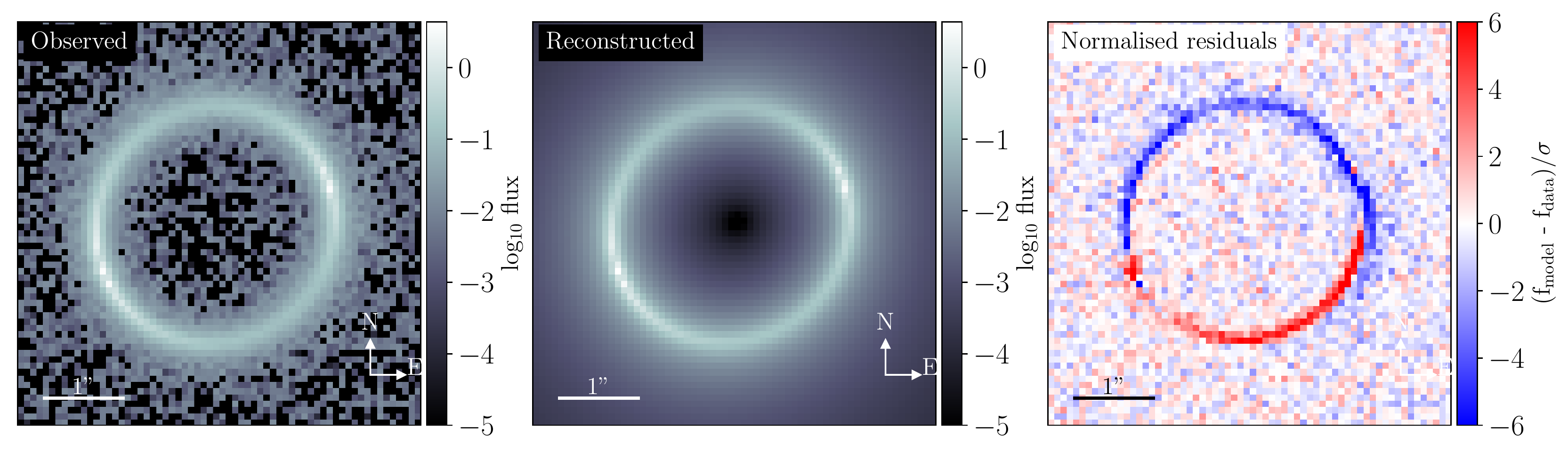}\\
    \includegraphics[width=\textwidth]{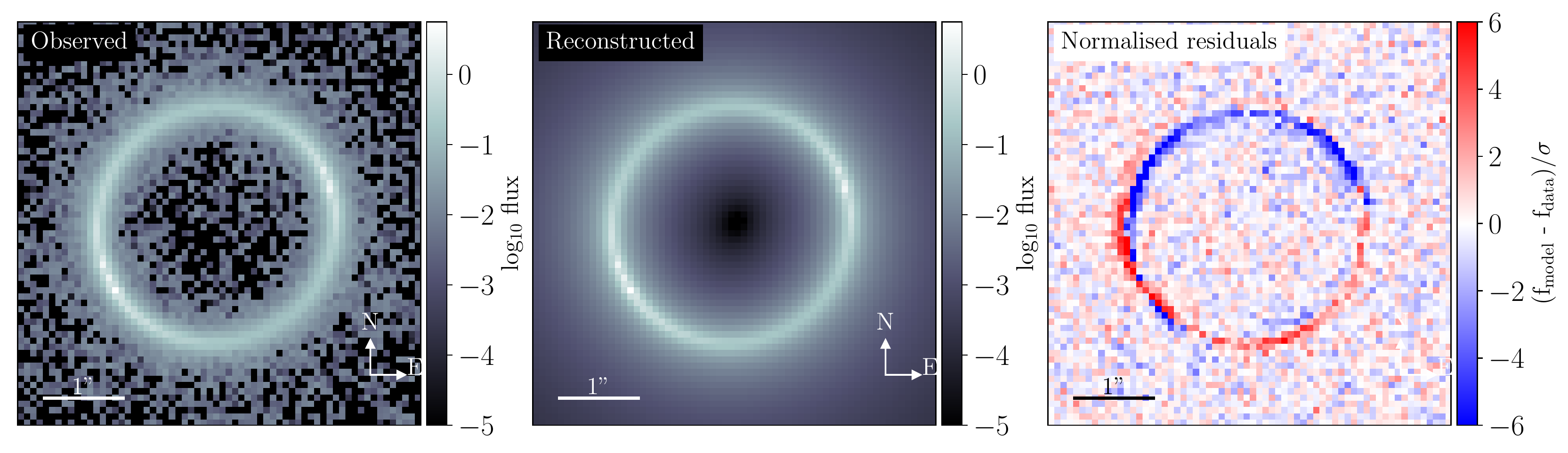}
    \caption{The image produced in the presence of the full population of haloes (top left) along with the reconstruction from the best-fit parameters found by the MCMC (top middle) and the normalised residuals of the difference between the two (top right). The bottom panels show the same but without the haloes~$h$ escaping the tidal regime ($\Delta_h \geq 0.1$).}
    \label{fig:dist_haloes_reconstruction}
\end{figure*}

In order to quantify the impact of beyond-tidal effects on the measurement of the LOS shear, we generate two images: the first one is the image that would be observed from a system consisting of our main lens plus $N$ haloes and the second one the image that would be observed from the same system but without the haloes~$h$ such that $\Delta_h \geq 0.1$. We do this using the multi-plane formalism in \lens. The two images are shown in the top and bottom left-hand panels in \cref{fig:dist_haloes_reconstruction}.

We then fit these two images using the \minmodel, plus an EPL mass profile for the main lens. For this part of the work, we changed our MCMC parameter inference method to use ensemble slice sampling \citep{Karamanis:2020zss} as implemented in the \texttt{zeus} package\footnote{\url{https://github.com/minaskar/zeus}.} \citep{Karamanis:2021tsx}, which we again modified \lens to include. We made this change due to the increased difficulty of the inference which required a faster and more powerful sampling method than the affine-invariant ensemble sampler provided in \texttt{emcee}.

\begin{figure*}
    \centering
    \includegraphics[width=0.49\textwidth]{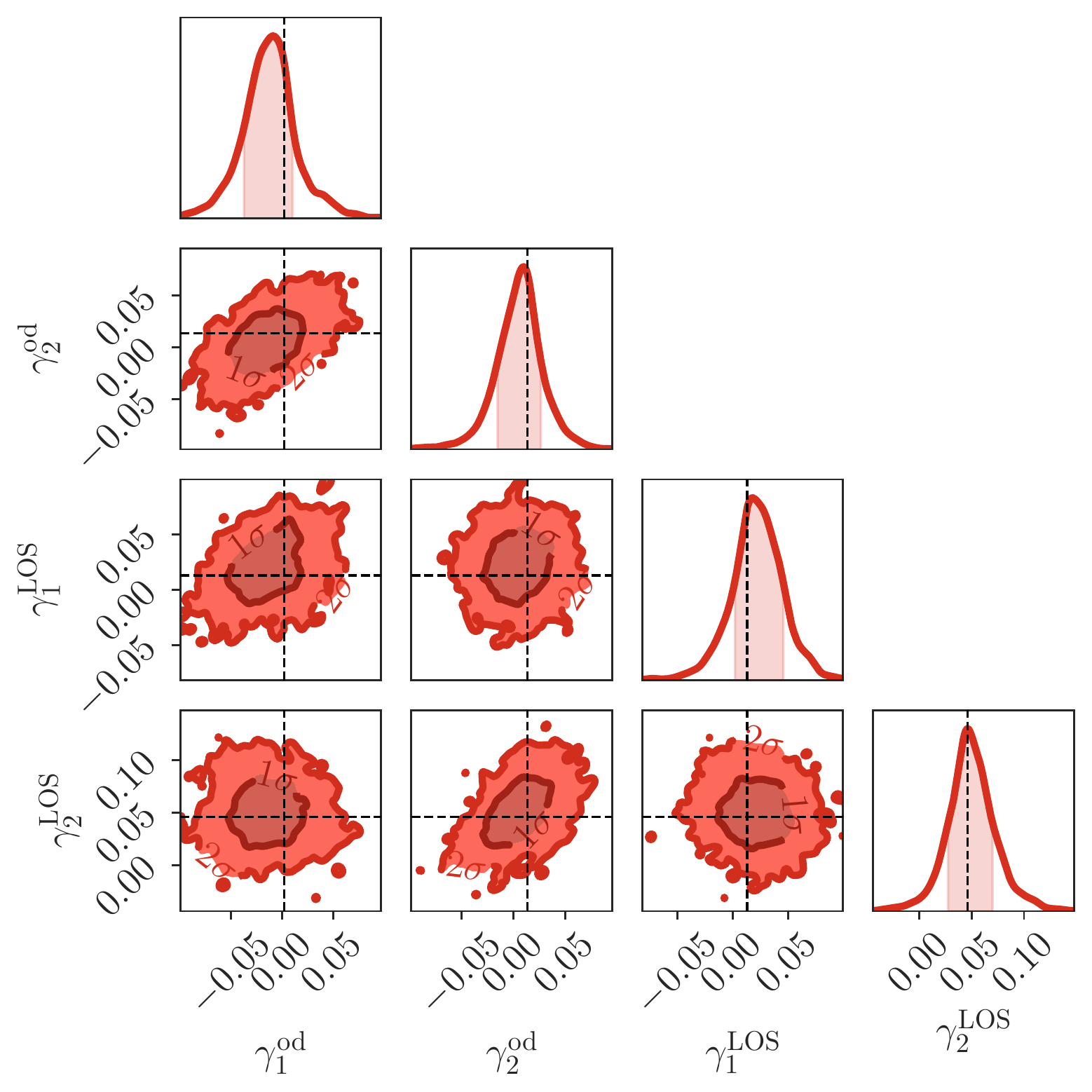}
    \hfill
    \includegraphics[width=0.49\textwidth]{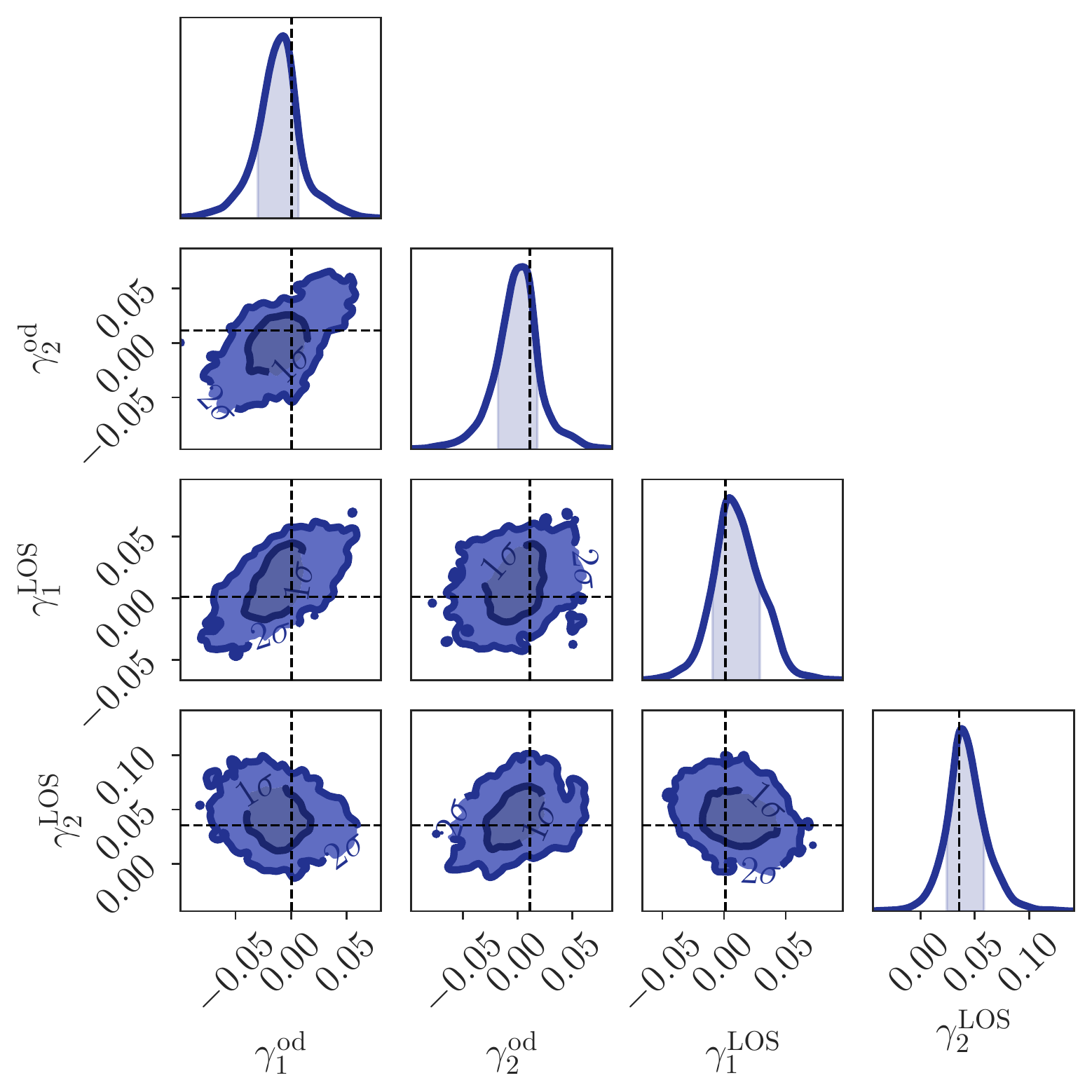}
    \caption{The one- and two-dimensional marginalised posterior distributions of the (od) and LOS shears that result from fitting the image produced by the total population of haloes (left) and the population without the haloes~$h$ for which $\Delta_h\geq 1$ (right). Dashed lines indicate the expected values on the optical axis, $\bm{\theta}=\bm{0}$.}
    \label{fig:dist_haloes_contours}
\end{figure*}

In \cref{fig:dist_haloes_contours}, we show the one- and two-dimensional marginalised posterior distributions of the (od) and LOS shears that result from fitting the images in \cref{fig:dist_haloes_reconstruction}. The dashed lines indicate expected values of $\gamma\e{od}(\bm{0})$ and $\gamma\e{LOS}(\bm{0}) = \gamma\e{os}(\bm{0}) + \gamma\e{od}(\bm{0}) - \gamma\e{ds}(\bm{0})$ exactly along the optical axis, in the absence of the main lens. We firstly consider the left-hand panel of this figure, which shows the results from fitting the image produced in the presence of the full population of haloes. The expected LOS shear is recovered with good accuracy, but with a precision degraded by about two orders of magnitude compared to the results of \cref{sec:advantage}, whose image was generated from a strictly tidal LOS perturbation model. Furthermore, as we can see from the top right panel of \cref{fig:dist_haloes_reconstruction}, there is an obvious strong residual between the observed image and the reconstructed image. \modification{The dipolar morphology of these residuals is reminiscent of the $\mathcal{F}$-type flexion signal predicted by \cite{Fleury:2021tke} (see Fig. 7 therein). Further investigation is nevertheless required in order to assess the detectability of such a signal in a realistic set-up.}

Taking now the contour plot in the right-hand panel of \cref{fig:dist_haloes_contours}, which results from fitting the image with the strongly non-tidal haloes removed, we can see that once again the LOS shear is recovered with good accuracy, while the precision is not significantly improved. The residuals exhibited in the case with the full halo population are still present \modification{with a similar amplitude}, as can be seen in the bottom right panel of \cref{fig:dist_haloes_reconstruction}.

These results indicate that, quite surprisingly, significant violations of the tidal regime do not entirely prevent one from measuring a notion of LOS shear from a strong lensing image. The tidal model is thus extremely resilient. Furthermore, the best-fit LOS shear appears to be unbiased with respect to the expected values on the optical axis.\footnote{\modification{This conclusion regarding the impact of higher-order LOS perturbations can be compared with the recent finding that the boxiness or diskiness of strong lenses do not significantly bias $H_0$ measurements~\citep{2022A&A...659A.127V}.}} Beyond-tidal effects then significantly increase the uncertainty on the shear measurement, which may be attributed to the fact that the shears are simply not single-valued across the image. They also produce visible patterns in the residuals of the best fit. A natural question is whether those patterns can be attributed to a combination of flexion signals; we shall address this point in future work.

We stress that the quantitative conclusions of this section should be taken with a pinch of salt. Indeed, in the scenario we constructed, haloes were initially randomly distributed in space without any clustering. However, in reality, halo positions are strongly correlated, with many small haloes found in the vicinity of larger ones, and following the cosmic two-point correlation function on large scales. Such correlations imply the presence of large coherent structures far from the LOS that could produce a rather uniform shear. In other words, we expect the role of beyond-tidal effects to be \emph{overestimated} in the present analysis. An important next step for future work will be to simulate a distribution of haloes that follows a cosmologically motivated two-point correlation function, possibly by obtaining a distribution of haloes and their physical characteristics such as mass and concentration from an $N$-body simulation. 

\section{Conclusions}
\label{sec:conclusion}

In this paper, we have provided a clear demonstration that the \minmodel, based on the dominant lens approximation and the tidal regime, to describe line-of-sight effects in strong gravitational lensing, evades the degeneracies inherent in other formalisms which try to address the same problem. Armed with this powerful model, we showed how the line-of-sight shear can be measured with percent-level accuracy from complicated strong lensing images even when source perturbations are not modelled. Furthermore, we tested the validity of the tidal approximation by simulating the effect of around $3.4\times 10^4$ dark matter haloes on a strong lensing image, finding that while the approximation may not be completely robust, LOS shear measurements are nevertheless possible.

Concretely, in \cref{sec:advantage}, we laid out the theoretical background to this work, beginning with the treatment of LOS effects in the tidal regime. We argued that the degeneracies present between the three shear terms in the lens equation in this regime means that they cannot be measured independently of each other or of the model parameters describing the main lens, but that a so-called \minmodel can be constructed in which these degeneracies are eliminated. Using \lens~-- modified for this work to include this LOS formalism -- we demonstrated the advantage that the \minmodel has over the full model, by constructing a simple lens model with shear and fitting the resulting image with both the full and \minmodel. As expected, the aforementioned degeneracies were present in the full model but not in the \minmodel.

Equipped with this demonstration of the advantage of the \minmodel, we applied it to the inference of the LOS shear from a set of 64 simulated strong lensing images, each produced by a rich composite lens model consisting of an elliptical dark matter halo offset from an elliptical baryonic core, along with LOS shear and perturbations to the source light, in \cref{sec:measurability_shear}. We fitted these images with a series of models, among which the optimal performance was obtained in the case where all the features of the simulated lenses are included save the foreground shear. This model is very similar to the composite lens models used to fit the H0LiCOW lens SDSS 1206+4332 by \cite{Birrer:2018vtm} and the STRIDES lens DES J0408-5354 by \cite{DES:2019fny}, both re-analysed by the TDCOSMO collaboration, again with a composite lens model \citep{Millon:2019slk}, and is similar to the multi-baryonic component lens models used by \cite{Williams:2020gud} and \cite{Nightingale:2019maw}, for example. In this case, we concluded that the LOS shear could be measured with an average absolute uncertainty of \modification{1\,\%}, without any systematic bias or degeneracy with the lens model parameters. We also showed how the inference of the LOS shear worsens with decreasing model complexity, as is to be expected. 


Finally, in \cref{sec:validity_tidal_regime}, we explored the validity of an important assumption: that LOS perturbers can be treated in the tidal regime. By simulating a scenario in which a volume around the LOS is populated with dark matter haloes, we analysed the images produced in the case where potentially beyond-tidal haloes were present and in the case where these haloes were removed. We observed a clear breakdown of the tidal regime, albeit with no effect on the accuracy of the measurement of the LOS shear. However, the precision was significantly worsened with respect to the best-case scenario of a very simple lens model in \cref{sec:advantage}. As a result of the way we modelled the distribution of the dark matter haloes, it is likely that we have overestimated their beyond-tidal effects, making this a very conservative estimate of the achievable precision of the LOS shear.

Lastly, we would like to highlight the important debate in the literature regarding whether external shear is truly \emph{external} or not -- in other words, if the distortions to a strong lensing image which have been traditionally attributed to the presence of perturbers external to the main lens are in fact due to another, perhaps poorly understood or controlled-for effect. One example of this was brought to light by \cite{VandeVyvere:2020koq}, who showed how incorrect truncation at the edges of a mass profile for a main lens can induce a spurious external shear. While this effect is important for reconstructions of lensing potentials from mass maps, it does not play a role in our current work due to our use of exact mass profiles which essentially extend to infinity. In that work, the spurious shear was also found to decrease to negligible magnitudes provided a large enough region for the mass map was used. The existence of external shear was recently further called into question by \cite{Etherington2023}, who showed how the best-fit value of an external shear parameter does not faithfully represent the actual shear if the lens mass is modelled by a simplistic elliptical power law. This finding is completely in agreement with the results of the current work.

In conclusion, the prospects for the use of the LOS shear as a new cosmological probe are excellent -- \emph{provided that the main lens mass is modelled correctly}. The LOS shear, which as we have shown is free from the usual degeneracies that plague typical strong lensing observables, will be able to be measured not only in current strong lensing data but from the many thousands of new strong lenses which are expected to be observed in the next five years by LSST and \textit{JWST}, and further afield, by \emph{Euclid}. The LOS shear thus has the potential to be employed in synergy with weak lensing surveys, magnifying the information that we receive from the inhomogeneous sky and boosting the constraints on cosmological parameters that are already in our hands.

\section*{Acknowledgements}
We thank all the participants of the LOS workshop held in Montpellier in June 2022 for the lively and enriching discussions about this work; we are particularly grateful to Simon Birrer for detailed discussions about \lens and to Daniel Johnson for the implementation of LOS effects in the Fermat potential which allowed for the time delay comparison in \Cref{appendix:lenstronomy}. We also thank Daniel Gilman for reviewing our \lens pull request, Anowar Shajib for providing information about TDCOSMO lens modelling, Sherry Suyu for suggesting that we consider a notion of signal-to-noise ratio to quantify the quality of our images, and Dominique Sluse and Liliya Williams for their detailed comments on the paper. We are also grateful to the anonymous referee for their comments which led to a substantial improvement of our results. Some of the colourmaps used in this work were provided by the \texttt{CMasher} package\footnote{\url{https://github.com/1313e/CMasher}.} \citep{vanderVelden2020}; other colour palettes were taken from \href{www.colorbrewer2.org}{\texttt{colorbrewer2}}. Some of the numerical computations for this work were done on the Sciama High Performance Compute (HPC) cluster, which is supported by the ICG, SEPNet and the University of Portsmouth. 
NBH is supported by a postdoctoral position funded by the French Commissariat à l’énergie atomique et aux énergies alternatives (CEA). JL's work was supported by a starting grant from the University of Montpellier and the French Ministry of Higher Education, Research and Innovation. MM acknowledges funding by the Agenzia Spaziale Italiana (\textsc{asi}) under agreement no. 2018-23-HH.0 and support from INFN/Euclid Sezione di Roma.

\section*{Data availability}
The software \lens was modified during this work, and the modifications are now part of the public version of that software. The full modification history of the code can be found in the pull request on GitHub: \url{https://github.com/lenstronomy/lenstronomy/pull/363}. Our wrapper for \lens built to produce and analyse the mock image catalogue presented here is publicly available at \url{https://github.com/nataliehogg/analosis}, along with the image data itself, while Jupyter notebooks for the rest of the analysis in this work can be found at \url{https://github.com/nataliehogg/los_proof_of_concept}.

\section*{CRedIT authorship contribution statement}

\textbf{Natalie B. Hogg:} Methodology, software, investigation, data curation, writing -- original draft, visualisation.
\textbf{Pierre Fleury:} Conceptualisation, methodology, software, validation, investigation, writing -- original draft, visualisation. 
\textbf{Julien Larena:} Conceptualisation, methodology, writing -- original draft.
\textbf{Matteo Martinelli:} Resources, writing -- review \& editing.

\bibliographystyle{mnras}
\bibliography{los}

\begin{thebibliography}{}
\makeatletter
\relax
\def\mn@urlcharsother{\let\do\@makeother \do\$\do\&\do\#\do\^\do\_\do\%\do\~}
\def\mn@doi{\begingroup\mn@urlcharsother \@ifnextchar [ {\mn@doi@}
  {\mn@doi@[]}}
\def\mn@doi@[#1]#2{\def\@tempa{#1}\ifx\@tempa\@empty \href
  {http://dx.doi.org/#2} {doi:#2}\else \href {http://dx.doi.org/#2} {#1}\fi
  \endgroup}
\def\mn@eprint#1#2{\mn@eprint@#1:#2::\@nil}
\def\mn@eprint@arXiv#1{\href {http://arxiv.org/abs/#1} {{\tt arXiv:#1}}}
\def\mn@eprint@dblp#1{\href {http://dblp.uni-trier.de/rec/bibtex/#1.xml}
  {dblp:#1}}
\def\mn@eprint@#1:#2:#3:#4\@nil{\def\@tempa {#1}\def\@tempb {#2}\def\@tempc
  {#3}\ifx \@tempc \@empty \let \@tempc \@tempb \let \@tempb \@tempa \fi \ifx
  \@tempb \@empty \def\@tempb {arXiv}\fi \@ifundefined
  {mn@eprint@\@tempb}{\@tempb:\@tempc}{\expandafter \expandafter \csname
  mn@eprint@\@tempb\endcsname \expandafter{\@tempc}}}

\bibitem[\protect\citeauthoryear{Aghanim et~al.}{Aghanim
  et~al.}{2020}]{Aghanim:2018eyx}
Aghanim N.,  et~al., 2020, \mn@doi [Astronomy \& Astrophysics]
  {10.1051/0004-6361/201833910}, 641, A6

\bibitem[\protect\citeauthoryear{{Astropy Collaboration} et~al.,}{{Astropy
  Collaboration} et~al.}{2022}]{astropy:2022}
{Astropy Collaboration} et~al., 2022, \mn@doi [The Astrophysical Journal]
  {10.3847/1538-4357/ac7c74}, \href
  {https://ui.adsabs.harvard.edu/abs/2022ApJ...935..167A} {935, 167}

\bibitem[\protect\citeauthoryear{Bacon, Goldberg, Rowe  \& Taylor}{Bacon
  et~al.}{2006}]{Bacon:2005qr}
Bacon D.~J.,  Goldberg D.~M.,  Rowe B. T.~P.,   Taylor A.~N.,  2006, \mn@doi
  [Monthly Notices of the Royal Astronomical Society]
  {10.1111/j.1365-2966.2005.09624.x}, 365, 414

\bibitem[\protect\citeauthoryear{{Bar-Kana}}{{Bar-Kana}}{1996}]{1996ApJ...468...17B}
{Bar-Kana} R.,  1996, \mn@doi [The Astrophysical Journal] {10.1086/177666},
  \href {https://ui.adsabs.harvard.edu/abs/1996ApJ...468...17B} {468, 17}

\bibitem[\protect\citeauthoryear{Birrer \& Amara}{Birrer \&
  Amara}{2018}]{Birrer:2018xgm}
Birrer S.,  Amara A.,  2018, \mn@doi [Physics of the Dark Universe]
  {https://doi.org/10.1016/j.dark.2018.11.002}, 22, 189

\bibitem[\protect\citeauthoryear{Birrer, Welschen, Amara  \& Refregier}{Birrer
  et~al.}{2017}]{Birrer:2016xku}
Birrer S.,  Welschen C.,  Amara A.,   Refregier A.,  2017, \mn@doi [Journal of
  Cosmology and Astroparticle Physics] {10.1088/1475-7516/2017/04/049}, 04, 049

\bibitem[\protect\citeauthoryear{Birrer, Refregier  \& Amara}{Birrer
  et~al.}{2018}]{Birrer:2017sge}
Birrer S.,  Refregier A.,   Amara A.,  2018, \mn@doi [The Astrophysical
  Journal] {10.3847/2041-8213/aaa1de}, 852, L14

\bibitem[\protect\citeauthoryear{Birrer et~al.}{Birrer
  et~al.}{2019}]{Birrer:2018vtm}
Birrer S.,  et~al., 2019, \mn@doi [Monthly Notices of the Royal Astronomical
  Society] {10.1093/mnras/stz200}, 484, 4726

\bibitem[\protect\citeauthoryear{Birrer et~al.,}{Birrer
  et~al.}{2021}]{Birrer2021}
Birrer S.,  et~al., 2021, \mn@doi [Journal of Open Source Software]
  {10.21105/joss.03283}, 6, 3283

\bibitem[\protect\citeauthoryear{{Blandford} \& {Narayan}}{{Blandford} \&
  {Narayan}}{1986}]{1986ApJ...310..568B}
{Blandford} R.,  {Narayan} R.,  1986, \mn@doi [The Astrophysical Journal]
  {10.1086/164709}, \href
  {https://ui.adsabs.harvard.edu/abs/1986ApJ...310..568B} {310, 568}

\bibitem[\protect\citeauthoryear{Despali, Vegetti, White, Giocoli  \& van~den
  Bosch}{Despali et~al.}{2018}]{Despali:2017ksx}
Despali G.,  Vegetti S.,  White S. D.~M.,  Giocoli C.,   van~den Bosch F.~C.,
  2018, \mn@doi [Monthly Notices of the Royal Astronomical Society]
  {10.1093/mnras/sty159}, 475, 5424

\bibitem[\protect\citeauthoryear{Dhanasingham, Cyr-Racine, Peter, Benson  \&
  Gilman}{Dhanasingham et~al.}{2022}]{Dhanasingham:2022nox}
Dhanasingham B.,  Cyr-Racine F.-Y.,  Peter A. H.~G.,  Benson A.,   Gilman D.,
  2022, {Interlopers speak out: Studying the dark universe using small-scale
  lensing anisotropies} (\mn@eprint {arXiv} {2203.13775})

\bibitem[\protect\citeauthoryear{Diemer}{Diemer}{2018}]{Diemer:2017bwl}
Diemer B.,  2018, \mn@doi [The Astrophysical Journal Supplement]
  {10.3847/1538-4365/aaee8c}, 239, 35

\bibitem[\protect\citeauthoryear{Diemer \& Joyce}{Diemer \&
  Joyce}{2019}]{Diemer:2018vmz}
Diemer B.,  Joyce M.,  2019, \mn@doi [The Astrophysical Journal]
  {10.3847/1538-4357/aafad6}, 871, 168

\bibitem[\protect\citeauthoryear{{Etherington} et~al.,}{{Etherington}
  et~al.}{2023}]{Etherington2023}
{Etherington} A.,  et~al., 2023, arXiv e-prints, \href
  {https://ui.adsabs.harvard.edu/abs/2023arXiv230105244E} {p. arXiv:2301.05244}

\bibitem[\protect\citeauthoryear{{Falco}, {Gorenstein}  \& {Shapiro}}{{Falco}
  et~al.}{1985}]{Falco1985}
{Falco} E.~E.,  {Gorenstein} M.~V.,   {Shapiro} I.~I.,  1985, \mn@doi [The
  Astrophysical Journal Letters] {10.1086/184422}, \href
  {https://ui.adsabs.harvard.edu/abs/1985ApJ...289L...1F} {289, L1}

\bibitem[\protect\citeauthoryear{Fleury, Larena  \& Uzan}{Fleury
  et~al.}{2021a}]{Fleury:2021tke}
Fleury P.,  Larena J.,   Uzan J.-P.,  2021a, \mn@doi [Journal of Cosmology and
  Astroparticle Physics] {10.1088/1475-7516/2021/08/024}, 08, 024

\bibitem[\protect\citeauthoryear{{Fleury}, {Larena}  \& {Uzan}}{{Fleury}
  et~al.}{2021b}]{2021CQGra..38h5002F}
{Fleury} P.,  {Larena} J.,   {Uzan} J.-P.,  2021b, \mn@doi [Classical and
  Quantum Gravity] {10.1088/1361-6382/abea2d}, \href
  {https://ui.adsabs.harvard.edu/abs/2021CQGra..38h5002F} {38, 085002}

\bibitem[\protect\citeauthoryear{{Foreman-Mackey}, {Hogg}, {Lang}  \&
  {Goodman}}{{Foreman-Mackey} et~al.}{2013}]{ForemanMackey2013}
{Foreman-Mackey} D.,  {Hogg} D.~W.,  {Lang} D.,   {Goodman} J.,  2013, \mn@doi
  [Publications of the Astronomical Society of the Pacific] {10.1086/670067},
  \href {https://ui.adsabs.harvard.edu/abs/2013PASP..125..306F} {125, 306}

\bibitem[\protect\citeauthoryear{Gilman, Birrer  \& Treu}{Gilman
  et~al.}{2020}]{Gilman_2020}
Gilman D.,  Birrer S.,   Treu T.,  2020, \mn@doi [Astronomy \& Astrophysics]
  {10.1051/0004-6361/202038829}, 642, A194

\bibitem[\protect\citeauthoryear{Gomer \& Williams}{Gomer \&
  Williams}{2021}]{Gomer:2021gio}
Gomer M.~R.,  Williams L. L.~R.,  2021, \mn@doi [Monthly Notices of the Royal
  Astronomical Society] {10.1093/mnras/stab930}, 504, 1340

\bibitem[\protect\citeauthoryear{{Hall}}{{Hall}}{2021}]{2021MNRAS.505.4935H}
{Hall} A.,  2021, \mn@doi [Monthly Notices of the Royal Astronomical Society]
  {10.1093/mnras/stab1563}, \href
  {https://ui.adsabs.harvard.edu/abs/2021MNRAS.505.4935H} {505, 4935}

\bibitem[\protect\citeauthoryear{He et~al.}{He et~al.}{2022}]{He:2021rjd}
He Q.,  et~al., 2022, \mn@doi [Monthly Notices of the Royal Astronomical
  Society] {10.1093/mnras/stac759}, 512, 5862

\bibitem[\protect\citeauthoryear{{Hinton}}{{Hinton}}{2016}]{Hinton2016}
{Hinton} S.~R.,  2016, \mn@doi [The Journal of Open Source Software]
  {10.21105/joss.00045}, \href
  {http://adsabs.harvard.edu/abs/2016JOSS....1...45H} {1, 00045}

\bibitem[\protect\citeauthoryear{Karamanis \& Beutler}{Karamanis \&
  Beutler}{2021}]{Karamanis:2020zss}
Karamanis M.,  Beutler F.,  2021, \mn@doi [Statistics and Computing]
  {10.1007/s11222-021-10038-2}, 31, 61

\bibitem[\protect\citeauthoryear{Karamanis, Beutler  \& Peacock}{Karamanis
  et~al.}{2021}]{Karamanis:2021tsx}
Karamanis M.,  Beutler F.,   Peacock J.~A.,  2021, \mn@doi [Monthly Notices of
  the Royal Astronomical Society] {10.1093/mnras/stab2867}, 508, 3589

\bibitem[\protect\citeauthoryear{{Kormann}, {Schneider}  \&
  {Bartelmann}}{{Kormann} et~al.}{1994}]{1994A&A...284..285K}
{Kormann} R.,  {Schneider} P.,   {Bartelmann} M.,  1994, Astronomy \&
  Astrophysics, \href {https://ui.adsabs.harvard.edu/abs/1994A&A...284..285K}
  {284, 285}

\bibitem[\protect\citeauthoryear{{Kovner}}{{Kovner}}{1987}]{1987ApJ...316...52K}
{Kovner} I.,  1987, \mn@doi [The Astrophysical Journal] {10.1086/165179}, \href
  {https://ui.adsabs.harvard.edu/abs/1987ApJ...316...52K} {316, 52}

\bibitem[\protect\citeauthoryear{Kuhlen, Guedes, Pillepich, Madau  \&
  Mayer}{Kuhlen et~al.}{2013}]{Kuhlen:2012qw}
Kuhlen M.,  Guedes J.,  Pillepich A.,  Madau P.,   Mayer L.,  2013, \mn@doi
  [The Astrophysical Journal] {10.1088/0004-637X/765/1/10}, 765, 10

\bibitem[\protect\citeauthoryear{Kuhn, Birrer, Bruderer, Amara  \&
  Refregier}{Kuhn et~al.}{2021}]{Kuhn:2020wpy}
Kuhn F.~A.,  Birrer S.,  Bruderer C.,  Amara A.,   Refregier A.,  2021, \mn@doi
  [Journal of Cosmology and Astroparticle Physics]
  {10.1088/1475-7516/2021/04/010}, 04, 010

\bibitem[\protect\citeauthoryear{Massey et~al.}{Massey
  et~al.}{2015}]{Massey:2015dkw}
Massey R.,  et~al., 2015, \mn@doi [Monthly Notices of the Royal Astronomical
  Society] {10.1093/mnras/stv467}, 449, 3393

\bibitem[\protect\citeauthoryear{McCully, Keeton, Wong  \& Zabludoff}{McCully
  et~al.}{2014}]{McCully:2013fga}
McCully C.,  Keeton C.~R.,  Wong K.~C.,   Zabludoff A.~I.,  2014, \mn@doi
  [Monthly Notices of the Royal Astronomical Society] {10.1093/mnras/stu1316},
  443, 3631

\bibitem[\protect\citeauthoryear{Millon et~al.}{Millon
  et~al.}{2020}]{Millon:2019slk}
Millon M.,  et~al., 2020, \mn@doi [Astronomy \& Astrophysics]
  {10.1051/0004-6361/201937351}, 639, A101

\bibitem[\protect\citeauthoryear{Navarro, Frenk  \& White}{Navarro
  et~al.}{1996}]{Navarro:1995iw}
Navarro J.~F.,  Frenk C.~S.,   White S. D.~M.,  1996, \mn@doi [The
  Astrophysical Journal] {10.1086/177173}, 462, 563

\bibitem[\protect\citeauthoryear{{Nightingale} \& {Dye}}{{Nightingale} \&
  {Dye}}{2015}]{Nightingale2015}
{Nightingale} J.~W.,  {Dye} S.,  2015, \mn@doi [Monthly Notices of the Royal
  Astronomical Society] {10.1093/mnras/stv1455}, \href
  {https://ui.adsabs.harvard.edu/abs/2015MNRAS.452.2940N} {452, 2940}

\bibitem[\protect\citeauthoryear{Nightingale, Dye  \& Massey}{Nightingale
  et~al.}{2018}]{Nightingale:2017cdh}
Nightingale J.,  Dye S.,   Massey R.,  2018, \mn@doi [Monthly Notices of the
  Royal Astronomical Society] {10.1093/mnras/sty1264}, 478, 4738

\bibitem[\protect\citeauthoryear{Nightingale, Massey, Harvey, Cooper,
  Etherington, Tam  \& Hayes}{Nightingale et~al.}{2019}]{Nightingale:2019maw}
Nightingale J.~W.,  Massey R.~J.,  Harvey D.~R.,  Cooper A.~P.,  Etherington
  A.,  Tam S.-I.,   Hayes R.~G.,  2019, \mn@doi [Monthly Notices of the Royal
  Astronomical Society] {10.1093/mnras/stz2220}, 489, 2049

\bibitem[\protect\citeauthoryear{Nightingale et~al.}{Nightingale
  et~al.}{2022}]{Nightingale:2022bhh}
Nightingale J.~W.,  et~al., 2022, {Scanning For Dark Matter Subhalos in Hubble
  Space Telescope Imaging of 54 Strong Lenses} (\mn@eprint {arXiv}
  {2209.10566})

\bibitem[\protect\citeauthoryear{Ostdiek, Diaz~Rivero  \& Dvorkin}{Ostdiek
  et~al.}{2022}]{Ostdiek:2020mvo}
Ostdiek B.,  Diaz~Rivero A.,   Dvorkin C.,  2022, \mn@doi [The Astrophysical
  Journal] {10.3847/1538-4357/ac2d8d}, 927, 83

\bibitem[\protect\citeauthoryear{Schneider}{Schneider}{1997}]{Schneider:1997bq}
Schneider P.,  1997, \mn@doi [Monthly Notices of the Royal Astronomical
  Society] {10.1093/mnras/292.3.673}, 292, 673

\bibitem[\protect\citeauthoryear{Schneider \& Sluse}{Schneider \&
  Sluse}{2014}]{Schneider:2013wga}
Schneider P.,  Sluse D.,  2014, \mn@doi [Astronomy \& Astrophysics]
  {10.1051/0004-6361/201322106}, 564, A103

\bibitem[\protect\citeauthoryear{{Seitz} \& {Schneider}}{{Seitz} \&
  {Schneider}}{1994}]{1994A&A...287..349S}
{Seitz} S.,  {Schneider} P.,  1994, Astronomy \& Astrophysics, \href
  {https://ui.adsabs.harvard.edu/abs/1994A&A...287..349S} {287, 349}

\bibitem[\protect\citeauthoryear{Sengül, Dvorkin, Ostdiek  \& Tsang}{Sengül
  et~al.}{2022}]{Seng_l_2022}
Sengül A.~{\c{C}}.,  Dvorkin C.,  Ostdiek B.,   Tsang A.,  2022, \mn@doi
  [Monthly Notices of the Royal Astronomical Society] {10.1093/mnras/stac1967},
  515, 4391

\bibitem[\protect\citeauthoryear{Shajib et~al.}{Shajib
  et~al.}{2020}]{DES:2019fny}
Shajib A.~J.,  et~al., 2020, \mn@doi [Monthly Notices of the Royal Astronomical
  Society] {10.1093/mnras/staa828}, 494, 6072

\bibitem[\protect\citeauthoryear{Suess, Kriek, Price  \& Barro}{Suess
  et~al.}{2019}]{Suess_2019}
Suess K.~A.,  Kriek M.,  Price S.~H.,   Barro G.,  2019, \mn@doi [The
  Astrophysical Journal] {10.3847/1538-4357/ab1bda}, 877, 103

\bibitem[\protect\citeauthoryear{{Tessore} \& {Metcalf}}{{Tessore} \&
  {Metcalf}}{2015}]{2015A&A...580A..79T}
{Tessore} N.,  {Metcalf} R.~B.,  2015, \mn@doi [Astronomy \& Astrophysics]
  {10.1051/0004-6361/201526773}, \href
  {https://ui.adsabs.harvard.edu/abs/2015A&A...580A..79T} {580, A79}

\bibitem[\protect\citeauthoryear{Tinker, Kravtsov, Klypin, Abazajian, Warren,
  Yepes, Gottlober  \& Holz}{Tinker et~al.}{2008}]{Tinker:2008ff}
Tinker J.~L.,  Kravtsov A.~V.,  Klypin A.,  Abazajian K.,  Warren M.~S.,  Yepes
  G.,  Gottlober S.,   Holz D.~E.,  2008, \mn@doi [The Astrophysical Journal]
  {10.1086/591439}, 688, 709

\bibitem[\protect\citeauthoryear{Van~de Vyvere, Sluse, Mukherjee, Xu  \&
  Birrer}{Van~de Vyvere et~al.}{2020}]{VandeVyvere:2020koq}
Van~de Vyvere L.,  Sluse D.,  Mukherjee S.,  Xu D.,   Birrer S.,  2020, \mn@doi
  [Astronomy \& Astrophysics] {10.1051/0004-6361/202038942}, 644, A108

\bibitem[\protect\citeauthoryear{{Van de Vyvere}, {Gomer}, {Sluse}, {Xu},
  {Birrer}, {Galan}  \& {Vernardos}}{{Van de Vyvere}
  et~al.}{2022}]{2022A&A...659A.127V}
{Van de Vyvere} L.,  {Gomer} M.~R.,  {Sluse} D.,  {Xu} D.,  {Birrer} S.,
  {Galan} A.,   {Vernardos} G.,  2022, \mn@doi [\aap]
  {10.1051/0004-6361/202141551}, \href
  {https://ui.adsabs.harvard.edu/abs/2022A&A...659A.127V} {659, A127}

\bibitem[\protect\citeauthoryear{Vegetti \& Koopmans}{Vegetti \&
  Koopmans}{2009}]{Vegetti_2009}
Vegetti S.,  Koopmans L. V.~E.,  2009, \mn@doi [Monthly Notices of the Royal
  Astronomical Society] {10.1111/j.1365-2966.2009.15559.x}, 400, 1583

\bibitem[\protect\citeauthoryear{Vegetti, Lagattuta, McKean, Auger, Fassnacht
  \& Koopmans}{Vegetti et~al.}{2012}]{Vegetti_2012}
Vegetti S.,  Lagattuta D.~J.,  McKean J.~P.,  Auger M.~W.,  Fassnacht C.~D.,
  Koopmans L. V.~E.,  2012, \mn@doi [Nature] {10.1038/nature10669}, 481, 341

\bibitem[\protect\citeauthoryear{Vegetti, Koopmans, Auger, Treu  \&
  Bolton}{Vegetti et~al.}{2014}]{Vegetti_2014}
Vegetti S.,  Koopmans L. V.~E.,  Auger M.~W.,  Treu T.,   Bolton A.~S.,  2014,
  \mn@doi [Monthly Notices of the Royal Astronomical Society]
  {10.1093/mnras/stu943}, 442, 2017

\bibitem[\protect\citeauthoryear{{\VAN{Velden}{Van}{van}}~der
  Velden}{{\VAN{Velden}{Van}{van}}~der Velden}{2020}]{vanderVelden2020}
{\VAN{Velden}{Van}{van}}~der Velden E.,  2020, \mn@doi [The Journal of Open
  Source Software] {10.21105/joss.02004}, \href
  {https://ui.adsabs.harvard.edu/abs/2020JOSS....5.2004V} {5, 2004}

\bibitem[\protect\citeauthoryear{Williams \& Zegeye}{Williams \&
  Zegeye}{2020}]{Williams:2020gud}
Williams L. L.~R.,  Zegeye D.,  2020, {Two-component mass models of the lensing
  galaxy in the quadruply imaged supernova iPTF16geu} (\mn@eprint {arXiv}
  {2006.09391}), \mn@doi{10.21105/astro.2006.09391}

\bibitem[\protect\citeauthoryear{{Windhorst} et~al.,}{{Windhorst}
  et~al.}{2011}]{Windhorst2011}
{Windhorst} R.~A.,  et~al., 2011, \mn@doi [The Astrophysical Journal
  Supplement] {10.1088/0067-0049/193/2/27}, \href
  {https://ui.adsabs.harvard.edu/abs/2011ApJS..193...27W} {193, 27}

\bibitem[\protect\citeauthoryear{Wong et~al.}{Wong et~al.}{2020}]{Wong:2019kwg}
Wong K.~C.,  et~al., 2020, \mn@doi [Monthly Notices of the Royal Astronomical
  Society] {10.1093/mnras/stz3094}, 498

\bibitem[\protect\citeauthoryear{Zhang, Mishra-Sharma  \& Dvorkin}{Zhang
  et~al.}{2022}]{Zhang:2022djp}
Zhang G.,  Mishra-Sharma S.,   Dvorkin C.,  2022, {Inferring subhalo effective
  density slopes from strong lensing observations with neural likelihood-ratio
  estimation} (\mn@eprint {arXiv} {2208.13796})

\makeatother
\end{thebibliography}

\appendix

\section{Modifying \lens} \label{appendix:lenstronomy}
The \lens software is a modular code designed primarily for modelling strong gravitational lenses, with numerous cosmological applications. The software contains a number of packages and subpackages which deal with the separate aspects of this task. For example, the \texttt{LensModel} package contains subpackages such as \texttt{Profiles}, which in turn contains a class for each lens mass profile which is available in \lens. 

The modifications to \lens made for this work follow this modular structure. We introduced a new subpackage under the \texttt{LensModel} package called \texttt{LineOfSight}, which contains a new class, \texttt{SinglePlaneLOS}. This new class contains functions to compute the  displacement angle, Hessian and Fermat potential in the LOS formalism. The class also inherits the ray shooting function (i.e. the computation of the source position $\bm{\beta}$), the lensing potential function and the functions for computing the lens mass enclosed in a given sphere or projected radius and lens mass density from the native \lens class \texttt{SinglePlane}, as these quantities are unchanged by LOS effects.

Furthermore, our new \texttt{LineOfSight} package contains a subpackage called \texttt{LOSModels}, inside which are the \texttt{LOS} and \texttt{LOS\_MINIMAL} classes. The functions encoding the LOS effects are defined in the \texttt{LOS} class, and these functions are then called in the \texttt{SinglePlaneLOS} class. The additional parameters associated with the LOS effects are also defined in the \texttt{LOS} class, and the parameters associated with the \minmodel are defined in \texttt{LOS\_MINIMAL}. The additional parameters are:
\begin{itemize}
    \item os: $\gamma_1^{\rm os}$, $\gamma_2^{\rm os}$, $\kappa^{\rm os}$, $\omega^{\rm os}$;
    \item od: $\gamma_1^{\rm od}$, $\gamma_2^{\rm od}$, $\kappa^{\rm od}$, $\omega^{\rm od}$;
    \item ds: $\gamma_1^{\rm ds}$, $\gamma_2^{\rm ds}$, $\kappa^{\rm ds}$, $\omega^{\rm ds}$;
\end{itemize}
in the \texttt{LOS} model and 
\begin{itemize}
    \item od: $\gamma_1^{\rm od}$, $\gamma_2^{\rm od}$, $\kappa^{\rm od}$, $\omega^{\rm od}$;
    \item LOS: $\gamma_1^{\rm LOS}$, $\gamma_2^{\rm LOS}$, $\kappa^{\rm LOS}$, $\omega^{\rm LOS}$;
\end{itemize}
in the \texttt{LOS\_MINIMAL} model. 

The final modifications we made to \lens were to introduce additional unit tests which ensure the validity of the functions and the quantities which they return each time the code is built. We created two types of unit test. The first checks that when all the additional LOS parameters are fixed to zero, the \texttt{SinglePlaneLOS} functions return the same values as the \texttt{SinglePlane} functions for a given source and lens. The second exploits the correspondence that exists between the LOS formalism and multi-plane lensing with three shear planes. This allows us to construct unit tests which check that the LOS modifications return the same values as those returned by the equivalent multi-plane set-up using the \texttt{MultiPlane} class of \lens.

For example, in \Cref{tab:unittests}, we show the output of the altered single-plane functions that we implemented in the  \texttt{SinglePlaneLOS} class with the equivalent \texttt{MultiPlane} settings for a simple lens model consisting of an elliptical power law lens model with three shear planes. The lens and shear parameters, image positions, and the lens and source redshifts were chosen at random to produce these results. 

The displacement angle computed by the \texttt{SinglePlaneLOS} class is identical to that computed in the multi-plane case to machine precision, $\Delta \alpha \sim 10^{-16}\,\mathrm{arcsec}$. The elements of the Hessian matrix in the LOS case are the same as in the multi-plane case at the level of at least $10^{-8}$. Finally, the time delay resulting from the modified Fermat potential in the \texttt{SinglePlaneLOS} class matches that of the multi-plane case at the level of $\mathcal{O}(10^{-11})$ days.

\begin{table*}
    \centering
    \caption{Example output of the \texttt{SinglePlaneLOS} and \texttt{MultiPlane} classes for comparison.}
    \begin{tabular}{Slccc}
    \hline     
    \hline 
    & \texttt{SinglePlaneLOS} &    \texttt{MultiPlane} & Difference \\
    \hline
     Displacement angle    & $\left(0.92460116,\ 0.77805976\right)''$  & $\left(0.92460116,\ 0.77805976\right)''$ & $\left(4.44089210\times 10^{-16},\ 3.33066907\times 10^{-16}\right)''$ \\
     Hessian & $\begin{bmatrix}
                0.11183298 & -0.0151521 \\
                -0.01004559 & 0.15331121
                \end{bmatrix}$ & 
                $\begin{bmatrix} 
                0.11183299 & -0.0151521 \\
                -0.01004559 & 0.15331119
                \end{bmatrix}$  &
                $\begin{bmatrix}
                6.50405486\times 10^{-9} & -3.40569642\times 10^{-10} \\
                -1.94965391\times 10^{-10}& -2.69573809\times 10^{-8}
                \end{bmatrix}$
                \\
     Time delay & $6.19293093$ days & $6.19293093$ days & $-1.18127730\times 10^{-11}$ days\\
     \hline
    \end{tabular}
    \label{tab:unittests}
\end{table*}

\section{Mock image catalogue model parameters} \label{appendix:composite_lens_parameters}
In this Appendix, we describe the full set of model parameters used to create the images in our mock catalogue. The numbers used to describe galaxy masses, half-light radii and \Sersic indices are freely inspired from \cite{Suess_2019}.

\subsection{Baryonic core: \Sersic ellipse}
This component is modelled using the \texttt{SERSIC\_ELLIPSE\_POTENTIAL} in \lens, which requires the angular half-light radius $R_{\rm \mSersic}$, the \Sersic index $n_{\rm \mSersic}$, the ellipticity components $e_1, e_2$, the effective convergence $k_{\rm eff}$ and the centre position $x,y$. 

We begin by drawing a redshift for each lens from a uniform distribution, $z \sim \mathcal{U}(0.4, 0.6)$ and from this compute the angular diameter distance to each lens, $D_{\rm od}$. Next, we draw the mass of the baryonic component randomly from a log-normal distribution, $\mathcal{M} \sim \ln \mathcal{N} \left[\ln(\mathcal{M}_{\rm mean}), \ln(2)/2\right]$, where the mean mass \modification{$\mathcal{M}_{\rm mean}= 2 \times 10^{11} \mathcal{M}_{\odot}$}. This distribution ensures that 95\% of the lens galaxies we simulate have a mass which is at most a factor of two smaller or larger than the mean mass. From this we compute the half-light radius of the baryonic core in arcseconds,
\begin{align}
\frac{R_{\rm \mSersic}}{\mathrm{arcsec}}
=
\frac{\mathcal{M}}{\mathcal{M}_{\rm mean}} \,
\frac{2\,\mathrm{kpc}}{D_{\rm od}}
\,
\frac{180 \times 3600}{\pi}  .
\end{align}
We also draw the values for the \Sersic index from a log-normal distribution, $n_{\rm \mSersic} \sim \ln \mathcal{N}[\ln(n_{\rm mean}), \ln(1.5)/2]$, where $n_{\rm mean} = 4$.

We compute the ellipticity components using the following expressions,
\begin{align}
    e_1 &= \frac{(1-q)}{(1+q)\cos(2\phi)},\\
    e_2 &= \frac{(1-q)}{(1+q)\sin(2\phi)},
\end{align}
where $q$ is the aspect ratio and $\phi$ the orientation angle of the ellipse. We draw the aspect ratio and orientations for each lens from uniform distributions, $q \sim \mathcal{U}(0.7, 1.0)$ and $\phi \sim \mathcal{U}(0, 2\pi)$ respectively.

We compute the effective convergence at the half-light radius,
\begin{align}
\kappa_{\rm eff} &= \frac{\mathcal{M}}{\Sigma_{\rm crit} I} \ ,
\end{align}
where $\mathcal{M}$ is the lens mass in units of $M_{\odot}$, $\Sigma_{\rm crit}$ is the critical surface density in the lens plane in units of $M_{\odot}/\text{arcsec}^2$ and the normalisation~$I$ is the integrated convergence for the given lens when $\kappa_{\rm eff} = 1$ in units of arcsec$^2$.

Finally, the centre of each lens is fixed to zero, $x=y=0$, i.e. exactly aligned with the optical axis.

\subsection{Lens light: \Sersic ellipse}
We model the lens light with an identical \Sersic ellipse profile so that the lens light traces the lens mass exactly. We use the \texttt{SERSIC\_ELLIPSE} light profile in \lens, which takes the same parameters as the \texttt{SERSIC\_ELLIPSE\_POTENTIAL} mass profile, less the effective convergence and with the addition of an apparent magnitude.

We start by computing the absolute magnitude of each lens,
\begin{align}
    M &= M_{\odot} - 2.5 \log_{10}\left(\frac{\mathcal{M}}{\Upsilon}\right),
\end{align}
where $M_{\odot} = 4.74$ is the absolute magnitude of the Sun and $\Upsilon = 2$ is the mass-to-light ratio for the lens galaxy. The apparent magnitude is then
\begin{align}
m = M + 5 \log_{10}\left(\frac{D}{1\,\mathrm{Mpc}}\right) + 25,
\end{align}
where $D$ is the luminosity distance to the lens, computed from the redshift and angular diameter distance, 
\begin{align}
    D = (1+z)^2 D_{\rm od}.
\end{align}
The remainder of the lens light parameters for each lens are the same as those used for the baryonic core.

\subsection{Dark matter halo: NFW ellipse}
We model the dark matter halo using the \texttt{NFW\_ELLIPSE} mass profile in \lens, which requires the scale radius, $R_{\rm s}$, the deflection angle at the scale radius $\alpha$, the ellipticity components $e_1, e_2$ and the centre position, $x,y$.

We start by drawing the ratio of dark matter to baryons in the lens galaxy from a uniform distribution, $\lambda \sim  \mathcal{U}(0.3, 1.0)$, and thus compute the mass of each halo via
\begin{equation}
    \mathcal{M}_{\rm halo} = \mathcal{M}/\lambda.
\end{equation}
We draw the concentration of each halo randomly from a log-normal distribution, \modification{$c \sim \ln \mathcal{N}(c_{\rm mean}, 0.2)$, where $c_{\rm mean} = 10^{1.1}$}. We then use \lens to compute the scale radius and deflection angle for each halo from the mass and concentration.

We follow the same procedure as for the ellipticity of the baryonic component to compute the ellipticity of each NFW halo. Lastly, we displace the centre of each halo from the baryonic core by drawing an offset (in parsecs) from a normal distribution, $o \sim \mathcal{N}(0, 300)$. \modification{This distribution is motivated by the findings of \cite{Kuhlen:2012qw}, who found a 300--400 pc offset in a simulation of a Milky Way-like galaxy. An even larger offset of 1.62 kpc has been observed in a central galaxy in the Abell 3827 cluster \citep{Massey:2015dkw}.}

\subsection{Source light: \Sersic ellipse}
As with the lens light, we model the source component using the \texttt{SERSIC\_ELLIPSE} light profile in \lens, and our methodology for computing the model parameters for each source is the same as followed for the lens light, except for two differences: the half-light radius and apparent magnitude are computed using the angular diameter distance from observer to source, $D_{\rm os}$, rather than observer to lens, $D_{\rm od}$; and the centre of the source, $x, y$, is allowed to be offset from the optical axis. We draw the source centre randomly from a uniform distribution on a disk, with the maximum allowed displacement from the optical axis being given by whichever is smallest: the half-light radius or the effective Einstein radius of the lens.

\modification{
\subsection{Einstein radius}

We estimate the Einstein radius of our composite lenses by neglecting the ellipticity and offset of their components, i.e. as if the lenses were axially symmetric; $\theta\e{E}$ is, thus, the solution of
\begin{equation}
\label{eq:estimate_Einstein_radius}
\theta\e{E} - \alpha\e{\mSersic}(\theta\e{E}) - \alpha\e{NFW}(\theta\e{E}) = 0 \ ,
\end{equation}
where $\alpha\e{\mSersic}$ and $\alpha\e{NFW}$ are the displacement angles individually produced by the baryon and dark-matter component of the lens. After drawing randomly their parameters, we compute $\theta\e{E}$ by solving \cref{eq:estimate_Einstein_radius} numerically, and we only keep the lenses such that $\theta\e{E} > 0.5''$, to ensure that the images we produce have arcs or rings which are reasonably well separated from the lens light.
}

\subsection{LOS shear}
We add the LOS shear to the images using the \texttt{LOS} model that we implemented in \lens. The two components of the od, os and ds shears, $\gamma_1$ and $\gamma_2$, are drawn from uniform distributions on a disk. This is done by drawing the square of each shear component randomly from a uniform distribution, $\gamma^2 \sim \mathcal{U}(0, \gamma_{\rm max}^2)$, where we set $\gamma_{\rm max} = 0.025$ to ensure that the LOS shear, $\gamma_{\rm LOS}$, is never larger than 5\% -- as we discussed in \Cref{sec:advantage}, large shears can induce rotation in the image with respect to the source.

The two components of each shear are then computed via
\begin{align}
    \gamma_1 &= \gamma \cos(2\phi), \\
    \gamma_2 &= \gamma \sin(2\phi),
\end{align}
where $\phi$ is drawn randomly from a uniform distribution, $\phi \sim \mathcal{U}(0, \pi)$.

\label{lastpage}
\end{document}